\documentclass[prd,amsmath,aps,floats,amssymb, floatfix, superscriptaddress,nofootinbib,twocolumn,preprintnumbers]{revtex4-2}

\usepackage{amsmath,amssymb,natbib,latexsym,times,subfigure}

\usepackage{lineno}

\usepackage[T1]{fontenc}
\usepackage{aecompl}

\usepackage{verbatim}
\usepackage{cprotect}
\usepackage{hhline,cuted}
\usepackage{ulem}

\usepackage{verbatim}
\usepackage[usenames,dvipsnames]{xcolor}
\usepackage{tabulary}
\usepackage{tabularx}
\usepackage{todonotes}
\usepackage{hyperref}

\newcommand\blfootnote[1]{%
  \begingroup
  \renewcommand\thefootnote{}\footnote{#1}%
  \addtocounter{footnote}{-1}%
  \endgroup
}

\newcommand\hmpcinv{\,h\,{\rm Mpc^{-1}}}

\allowdisplaybreaks
\begin{document}

\title{Constraints on dark matter to dark radiation conversion in the late universe with DES-Y1 and external data}
\author{A.~Chen $^\dagger$} \blfootnote{$\dagger$ Corresponding author: anqich@umich.edu}
\affiliation{Department of Physics, University of Michigan, Ann Arbor, MI 48109, USA}
\author{D.~Huterer}
\affiliation{Department of Physics, University of Michigan, Ann Arbor, MI 48109, USA}
\author{S.~Lee}
\affiliation{Department of Physics, Duke University Durham, NC 27708, USA}
\author{A.~Fert\'e}
\affiliation{Jet Propulsion Laboratory, California Institute of Technology, 4800 Oak Grove Dr., Pasadena, CA 91109, USA}
\author{N.~Weaverdyck}
\affiliation{Department of Physics, University of Michigan, Ann Arbor, MI 48109, USA}
\author{O.~Alves}
\affiliation{Instituto de F\'{i}sica Te\'orica, Universidade Estadual Paulista, S\~ao Paulo, Brazil}
\author{C.~D.~Leonard}
\affiliation{School of Mathematics, Statistics and Physics, Newcastle Unviersity, NE1 7RU, United Kingdom}
\author{N.~MacCrann}
\affiliation{Center for Cosmology and Astro-Particle Physics, The Ohio State University, Columbus, OH 43210, USA}
\affiliation{Department of Physics, The Ohio State University, Columbus, OH 43210, USA}
\author{M.~Raveri}
\affiliation{Kavli Institute for Cosmological Physics, University of Chicago, Chicago, IL 60637, USA}
\author{A.~Porredon}
\affiliation{Center for Cosmology and Astro-Particle Physics, The Ohio State University, Columbus, OH 43210, USA}
\affiliation{Institut d'Estudis Espacials de Catalunya (IEEC), 08034 Barcelona, Spain}
\affiliation{Institute of Space Sciences (ICE, CSIC),  Campus UAB, Carrer de Can Magrans, s/n,  08193 Barcelona, Spain}
\author{E.~Di Valentino}
\affiliation{Jodrell Bank Center for Astrophysics, School of Physics and Astronomy, University of Manchester, Oxford Road, Manchester, M13 9PL, UK}
\author{J.~Muir}
\affiliation{Kavli Institute for Particle Astrophysics \& Cosmology, P. O. Box 2450, Stanford University, Stanford, CA 94305, USA}
\author{P.~Lemos}
\affiliation{Department of Physics \& Astronomy, University College London, Gower Street, London, WC1E 6BT, UK}
\author{A.~R.~Liddle}
\affiliation{Instituto de Astrof\'{\i}sica e Ci\^{e}ncias do Espa\c{c}o, Faculdade de Ci\^{e}ncias, Universidade de Lisboa, 1769-016 Lisboa, Portugal}
\affiliation{Institute for Astronomy, University of Edinburgh, Edinburgh EH9 3HJ, UK}
\affiliation{Perimeter Institute for Theoretical Physics, 31 Caroline St. North, Waterloo, ON N2L 2Y5, Canada}
\author{J.~Blazek}
\affiliation{Center for Cosmology and Astro-Particle Physics, The Ohio State University, Columbus, OH 43210, USA}
\affiliation{Institute of Physics, Laboratory of Astrophysics, \'Ecole Polytechnique F\'ed\'erale de Lausanne (EPFL), Observatoire de Sauverny, 1290 Versoix, Switzerland}
\author{A.~Campos}
\affiliation{Department of Physics, Carnegie Mellon University, Pittsburgh, Pennsylvania 15312, USA}
\affiliation{Instituto de F\'{i}sica Te\'orica, Universidade Estadual Paulista, S\~ao Paulo, Brazil}
\author{R.~Cawthon}
\affiliation{Physics Department, 2320 Chamberlin Hall, University of Wisconsin-Madison, 1150 University Avenue Madison, WI  53706-1390}
\author{A.~Choi}
\affiliation{Center for Cosmology and Astro-Particle Physics, The Ohio State University, Columbus, OH 43210, USA}
\author{S.~Dodelson}
\affiliation{Department of Physics, Carnegie Mellon University, Pittsburgh, Pennsylvania 15312, USA}
\author{J.~Elvin-Poole}
\affiliation{Center for Cosmology and Astro-Particle Physics, The Ohio State University, Columbus, OH 43210, USA}
\affiliation{Department of Physics, The Ohio State University, Columbus, OH 43210, USA}
\author{D.~Gruen}
\affiliation{Department of Physics, Stanford University, 382 Via Pueblo Mall, Stanford, CA 94305, USA}
\affiliation{Kavli Institute for Particle Astrophysics \& Cosmology, P. O. Box 2450, Stanford University, Stanford, CA 94305, USA}
\affiliation{SLAC National Accelerator Laboratory, Menlo Park, CA 94025, USA}
\author{A.~J.~Ross}
\affiliation{Center for Cosmology and Astro-Particle Physics, The Ohio State University, Columbus, OH 43210, USA}
\author{L.~F.~Secco}
\affiliation{Department of Physics and Astronomy, University of Pennsylvania, Philadelphia, PA 19104, USA}
\author{I.~Sevilla-Noarbe}
\affiliation{Centro de Investigaciones Energ\'eticas, Medioambientales y Tecnol\'ogicas (CIEMAT), Madrid, Spain}
\author{E.~Sheldon}
\affiliation{Brookhaven National Laboratory, Bldg 510, Upton, NY 11973, USA}
\author{M.~A.~Troxel}
\affiliation{Department of Physics, Duke University Durham, NC 27708, USA}
\author{J.~Zuntz}
\affiliation{Institute for Astronomy, University of Edinburgh, Edinburgh EH9 3HJ, UK}
\author{T.~M.~C.~Abbott}
\affiliation{Cerro Tololo Inter-American Observatory, NSF's National Optical-Infrared Astronomy Research Laboratory, Casilla 603, La Serena, Chile}
\author{M.~Aguena}
\affiliation{Departamento de F\'isica Matem\'atica, Instituto de F\'isica, Universidade de S\~ao Paulo, CP 66318, S\~ao Paulo, SP, 05314-970, Brazil}
\affiliation{Laborat\'orio Interinstitucional de e-Astronomia - LIneA, Rua Gal. Jos\'e Cristino 77, Rio de Janeiro, RJ - 20921-400, Brazil}
\author{S.~Allam}
\affiliation{Fermi National Accelerator Laboratory, P. O. Box 500, Batavia, IL 60510, USA}
\author{J.~Annis}
\affiliation{Fermi National Accelerator Laboratory, P. O. Box 500, Batavia, IL 60510, USA}
\author{S.~Avila}
\affiliation{Instituto de Fisica Teorica UAM/CSIC, Universidad Autonoma de Madrid, 28049 Madrid, Spain}
\author{E.~Bertin}
\affiliation{CNRS, UMR 7095, Institut d'Astrophysique de Paris, F-75014, Paris, France}
\affiliation{Sorbonne Universit\'es, UPMC Univ Paris 06, UMR 7095, Institut d'Astrophysique de Paris, F-75014, Paris, France}
\author{S.~Bhargava}
\affiliation{Department of Physics and Astronomy, Pevensey Building, University of Sussex, Brighton, BN1 9QH, UK}
\author{S.~L.~Bridle}
\affiliation{Jodrell Bank Center for Astrophysics, School of Physics and Astronomy, University of Manchester, Oxford Road, Manchester, M13 9PL, UK}
\author{D.~Brooks}
\affiliation{Department of Physics \& Astronomy, University College London, Gower Street, London, WC1E 6BT, UK}
\author{A.~Carnero~Rosell}
\affiliation{Instituto de Astrofisica de Canarias, E-38205 La Laguna, Tenerife, Spain}
\affiliation{Universidad de La Laguna, Dpto. Astrof\'{i}sica, E-38206 La Laguna, Tenerife, Spain}
\author{M.~Carrasco~Kind}
\affiliation{Department of Astronomy, University of Illinois at Urbana-Champaign, 1002 W. Green Street, Urbana, IL 61801, USA}
\affiliation{National Center for Supercomputing Applications, 1205 West Clark St., Urbana, IL 61801, USA}
\author{J.~Carretero}
\affiliation{Institut de F\'{\i}sica d'Altes Energies (IFAE), The Barcelona Institute of Science and Technology, Campus UAB, 08193 Bellaterra (Barcelona) Spain}
\author{M.~Costanzi}
\affiliation{INAF-Osservatorio Astronomico di Trieste, via G. B. Tiepolo 11, I-34143 Trieste, Italy}
\affiliation{Institute for Fundamental Physics of the Universe, Via Beirut 2, 34014 Trieste, Italy}
\author{M.~Crocce}
\affiliation{Institut d'Estudis Espacials de Catalunya (IEEC), 08034 Barcelona, Spain}
\affiliation{Institute of Space Sciences (ICE, CSIC),  Campus UAB, Carrer de Can Magrans, s/n,  08193 Barcelona, Spain}
\author{L.~N.~da Costa}
\affiliation{Laborat\'orio Interinstitucional de e-Astronomia - LIneA, Rua Gal. Jos\'e Cristino 77, Rio de Janeiro, RJ - 20921-400, Brazil}
\affiliation{Observat\'orio Nacional, Rua Gal. Jos\'e Cristino 77, Rio de Janeiro, RJ - 20921-400, Brazil}
\author{M.~E.~S.~Pereira}
\affiliation{Department of Physics, University of Michigan, Ann Arbor, MI 48109, USA}
\author{T.~M.~Davis}
\affiliation{School of Mathematics and Physics, University of Queensland,  Brisbane, QLD 4072, Australia}
\author{P.~Doel}
\affiliation{Department of Physics \& Astronomy, University College London, Gower Street, London, WC1E 6BT, UK}
\author{T.~F.~Eifler}
\affiliation{Department of Astronomy/Steward Observatory, University of Arizona, 933 North Cherry Avenue, Tucson, AZ 85721-0065, USA}
\affiliation{Jet Propulsion Laboratory, California Institute of Technology, 4800 Oak Grove Dr., Pasadena, CA 91109, USA}
\author{I.~Ferrero}
\affiliation{Institute of Theoretical Astrophysics, University of Oslo. P.O. Box 1029 Blindern, NO-0315 Oslo, Norway}
\author{P.~Fosalba}
\affiliation{Institut d'Estudis Espacials de Catalunya (IEEC), 08034 Barcelona, Spain}
\affiliation{Institute of Space Sciences (ICE, CSIC),  Campus UAB, Carrer de Can Magrans, s/n,  08193 Barcelona, Spain}
\author{J.~Frieman}
\affiliation{Fermi National Accelerator Laboratory, P. O. Box 500, Batavia, IL 60510, USA}
\affiliation{Kavli Institute for Cosmological Physics, University of Chicago, Chicago, IL 60637, USA}
\author{J.~Garc\'ia-Bellido}
\affiliation{Instituto de Fisica Teorica UAM/CSIC, Universidad Autonoma de Madrid, 28049 Madrid, Spain}
\author{E.~Gaztanaga}
\affiliation{Institut d'Estudis Espacials de Catalunya (IEEC), 08034 Barcelona, Spain}
\affiliation{Institute of Space Sciences (ICE, CSIC),  Campus UAB, Carrer de Can Magrans, s/n,  08193 Barcelona, Spain}
\author{D.~W.~Gerdes}
\affiliation{Department of Astronomy, University of Michigan, Ann Arbor, MI 48109, USA}
\affiliation{Department of Physics, University of Michigan, Ann Arbor, MI 48109, USA}
\author{R.~A.~Gruendl}
\affiliation{Department of Astronomy, University of Illinois at Urbana-Champaign, 1002 W. Green Street, Urbana, IL 61801, USA}
\affiliation{National Center for Supercomputing Applications, 1205 West Clark St., Urbana, IL 61801, USA}
\author{J.~Gschwend}
\affiliation{Laborat\'orio Interinstitucional de e-Astronomia - LIneA, Rua Gal. Jos\'e Cristino 77, Rio de Janeiro, RJ - 20921-400, Brazil}
\affiliation{Observat\'orio Nacional, Rua Gal. Jos\'e Cristino 77, Rio de Janeiro, RJ - 20921-400, Brazil}
\author{G.~Gutierrez}
\affiliation{Fermi National Accelerator Laboratory, P. O. Box 500, Batavia, IL 60510, USA}
\author{S.~R.~Hinton}
\affiliation{School of Mathematics and Physics, University of Queensland,  Brisbane, QLD 4072, Australia}
\author{D.~L.~Hollowood}
\affiliation{Santa Cruz Institute for Particle Physics, Santa Cruz, CA 95064, USA}
\author{K.~Honscheid}
\affiliation{Center for Cosmology and Astro-Particle Physics, The Ohio State University, Columbus, OH 43210, USA}
\affiliation{Department of Physics, The Ohio State University, Columbus, OH 43210, USA}
\author{B.~Hoyle}
\affiliation{Faculty of Physics, Ludwig-Maximilians-Universit\"at, Scheinerstr. 1, 81679 Munich, Germany}
\affiliation{Max Planck Institute for Extraterrestrial Physics, Giessenbachstrasse, 85748 Garching, Germany}
\affiliation{Universit\"ats-Sternwarte, Fakult\"at f\"ur Physik, Ludwig-Maximilians Universit\"at M\"unchen, Scheinerstr. 1, 81679 M\"unchen, Germany}
\author{D.~J.~James}
\affiliation{Center for Astrophysics $\vert$ Harvard \& Smithsonian, 60 Garden Street, Cambridge, MA 02138, USA}
\author{M.~Jarvis}
\affiliation{Department of Physics and Astronomy, University of Pennsylvania, Philadelphia, PA 19104, USA}
\author{K.~Kuehn}
\affiliation{Australian Astronomical Optics, Macquarie University, North Ryde, NSW 2113, Australia}
\affiliation{Lowell Observatory, 1400 Mars Hill Rd, Flagstaff, AZ 86001, USA}
\author{O.~Lahav}
\affiliation{Department of Physics \& Astronomy, University College London, Gower Street, London, WC1E 6BT, UK}
\author{M.~A.~G.~Maia}
\affiliation{Laborat\'orio Interinstitucional de e-Astronomia - LIneA, Rua Gal. Jos\'e Cristino 77, Rio de Janeiro, RJ - 20921-400, Brazil}
\affiliation{Observat\'orio Nacional, Rua Gal. Jos\'e Cristino 77, Rio de Janeiro, RJ - 20921-400, Brazil}
\author{J.~L.~Marshall}
\affiliation{George P. and Cynthia Woods Mitchell Institute for Fundamental Physics and Astronomy, and Department of Physics and Astronomy, Texas A\&M University, College Station, TX 77843,  USA}
\author{F.~Menanteau}
\affiliation{Department of Astronomy, University of Illinois at Urbana-Champaign, 1002 W. Green Street, Urbana, IL 61801, USA}
\affiliation{National Center for Supercomputing Applications, 1205 West Clark St., Urbana, IL 61801, USA}
\author{R.~Miquel}
\affiliation{Instituci\'o Catalana de Recerca i Estudis Avan\c{c}ats, E-08010 Barcelona, Spain}
\affiliation{Institut de F\'{\i}sica d'Altes Energies (IFAE), The Barcelona Institute of Science and Technology, Campus UAB, 08193 Bellaterra (Barcelona) Spain}
\author{R.~Morgan}
\affiliation{Physics Department, 2320 Chamberlin Hall, University of Wisconsin-Madison, 1150 University Avenue Madison, WI  53706-1390}
\author{A.~Palmese}
\affiliation{Fermi National Accelerator Laboratory, P. O. Box 500, Batavia, IL 60510, USA}
\affiliation{Kavli Institute for Cosmological Physics, University of Chicago, Chicago, IL 60637, USA}
\author{F.~Paz-Chinch\'{o}n}
\affiliation{Institute of Astronomy, University of Cambridge, Madingley Road, Cambridge CB3 0HA, UK}
\affiliation{National Center for Supercomputing Applications, 1205 West Clark St., Urbana, IL 61801, USA}
\author{A.~A.~Plazas}
\affiliation{Department of Astrophysical Sciences, Princeton University, Peyton Hall, Princeton, NJ 08544, USA}
\author{A.~Roodman}
\affiliation{Kavli Institute for Particle Astrophysics \& Cosmology, P. O. Box 2450, Stanford University, Stanford, CA 94305, USA}
\affiliation{SLAC National Accelerator Laboratory, Menlo Park, CA 94025, USA}
\author{E.~Sanchez}
\affiliation{Centro de Investigaciones Energ\'eticas, Medioambientales y Tecnol\'ogicas (CIEMAT), Madrid, Spain}
\author{V.~Scarpine}
\affiliation{Fermi National Accelerator Laboratory, P. O. Box 500, Batavia, IL 60510, USA}
\author{M.~Schubnell}
\affiliation{Department of Physics, University of Michigan, Ann Arbor, MI 48109, USA}
\author{S.~Serrano}
\affiliation{Institut d'Estudis Espacials de Catalunya (IEEC), 08034 Barcelona, Spain}
\affiliation{Institute of Space Sciences (ICE, CSIC),  Campus UAB, Carrer de Can Magrans, s/n,  08193 Barcelona, Spain}
\author{M.~Smith}
\affiliation{School of Physics and Astronomy, University of Southampton,  Southampton, SO17 1BJ, UK}
\author{E.~Suchyta}
\affiliation{Computer Science and Mathematics Division, Oak Ridge National Laboratory, Oak Ridge, TN 37831}
\author{G.~Tarle}
\affiliation{Department of Physics, University of Michigan, Ann Arbor, MI 48109, USA}
\author{D.~Thomas}
\affiliation{Institute of Cosmology and Gravitation, University of Portsmouth, Portsmouth, PO1 3FX, UK}
\author{C.~To}
\affiliation{Department of Physics, Stanford University, 382 Via Pueblo Mall, Stanford, CA 94305, USA}
\affiliation{Kavli Institute for Particle Astrophysics \& Cosmology, P. O. Box 2450, Stanford University, Stanford, CA 94305, USA}
\affiliation{SLAC National Accelerator Laboratory, Menlo Park, CA 94025, USA}
\author{T.~N.~Varga}
\affiliation{Max Planck Institute for Extraterrestrial Physics, Giessenbachstrasse, 85748 Garching, Germany}
\affiliation{Universit\"ats-Sternwarte, Fakult\"at f\"ur Physik, Ludwig-Maximilians Universit\"at M\"unchen, Scheinerstr. 1, 81679 M\"unchen, Germany}
\author{J.~Weller}
\affiliation{Max Planck Institute for Extraterrestrial Physics, Giessenbachstrasse, 85748 Garching, Germany}
\affiliation{Universit\"ats-Sternwarte, Fakult\"at f\"ur Physik, Ludwig-Maximilians Universit\"at M\"unchen, Scheinerstr. 1, 81679 M\"unchen, Germany}
\author{R.D.~Wilkinson}
\affiliation{Department of Physics and Astronomy, Pevensey Building, University of Sussex, Brighton, BN1 9QH, UK}

\collaboration{DES Collaboration}

\date{\today}

\label{firstpage}
\begin{abstract}
We study a phenomenological class of models where dark matter converts to dark radiation in the low redshift epoch. This class of models, dubbed DMDR, characterizes the evolution of comoving dark matter density with two extra parameters, and may be able to help alleviate the observed discrepancies between early- and late-time probes of the universe.
We investigate how the  conversion affects key cosmological observables such as the CMB temperature and matter power spectra. Combining 3x2pt data from Year 1 of the Dark Energy Survey, {\it Planck}-2018 CMB temperature and polarization data, supernovae (SN) Type Ia data from Pantheon, and baryon acoustic oscillation (BAO) data from BOSS DR12, MGS and 6dFGS, we place new constraints on the amount of dark matter that has converted to dark radiation and the rate of this conversion. The fraction of the dark matter that has converted since the beginning of the universe in units of the current amount of dark matter, $\zeta$, is constrained at 68\% confidence level to be $<0.32$ for DES-Y1 3x2pt data, $<0.030$  for CMB+SN+BAO data, and $<0.037$ for the combined dataset. The probability that the DES and CMB+SN+BAO datasets are concordant increases from 4\% for the $\Lambda$CDM model to 8\% (less tension) for DMDR. The tension in $S_8 = \sigma_8 \sqrt{\Omega_{\rm m}/0.3}$ between DES-Y1 3x2pt and CMB+SN+BAO is slightly reduced from $2.3\sigma$ to $1.9\sigma$. We find no reduction in the Hubble tension when the combined data is compared to distance-ladder measurements in the DMDR model. The maximum-posterior goodness-of-fit  statistics of DMDR and $\Lambda$CDM model are comparable, indicating no preference for the DMDR cosmology over $\Lambda$CDM.

\end{abstract}


\maketitle

\section{Introduction}

Over the past few years, there has been a notable improvement in both the variety and precision of cosmological probes. Signals predicted long ago, such as gravitational waves and global 21-cm absorption, were finally observed, providing new insights and solidifying our understanding of the universe. The enhanced precision of  relatively mature observational techniques such as measurements of galaxy clustering, weak lensing, and anisotropies in the cosmic microwave background (CMB) temperature and polarization fields has allowed us to test the $\Lambda$CDM paradigm to an unprecedented degree.  

Recent cosmological observations have revealed a discrepancy in the inferred Hubble constant at $\gtrsim 4 \sigma$  level between early- and late-universe probes \cite{planck2018,verdetensions,Riess:2020sih}. With a strengthening of the various steps in the local distance-ladder measurements of $H_0$, as well as tightening constraints of medium-to-high redshift probes such as strong and weak gravitational lensing, the Hubble tension is becoming more significant \cite{huang2018near,ariess,wendy,h0licow} and enormous effort has been devoted to understanding its origin.  A number of theories have thus far been proposed to help ameliorate or resolve the tension \cite{DiValentino:2019ffd,poulin2019early,lin2019acoustic,blinov2020warm,sola2020brans,jedamzik2020relieving,elizalde2020analysis,li2020generalised,Yang:2020myd,hart2020updated, ballardini2020scalar}, but so far none have done so to a satisfactory degree.

A parallel development over the last few years has been the consistently lower value of the amplitude of mass fluctuations $\sigma_8$ measured in gravitational lensing compared to that measured by the CMB experiments \cite{kids2016,leauthaud2017lensing,lin2017cosmological,desy1,DiValentino:2018gcu,2019kidscosebi}. While not currently statistically as strong as the Hubble tension, the persistence of the $\sigma_8$ measurement discrepancies, as well as their possible origin as a mismatch between the geometrical measures and the growth of structure expected in the currently-dominant $\Lambda$CDM paradigm, deserves special attention. It would be very exciting, and compelling, if both the $H_0$ and $\sigma_8$ tensions were solved simultaneously, though the success of extant models on this front is at best mixed \cite{buen2015non, murgia2016constraints,di2018reducing,hill2020early,ivanov2020constraining,klypin2020clustering}.

One possible explanation for why weak lensing surveys measure a smaller amplitude of fluctuations than the CMB is that the present-day matter content has decreased at a higher rate than predicted by $\Lambda$CDM model.
Models where dark matter converts into a new species with radiation properties that is not directly detectable (hence `dark radiation') can enable such a trend. These models also have the potential to reconcile the Hubble tension, as they predict a smaller matter content as time evolves. Accordingly, dark energy dominates faster than in $\Lambda$CDM in these models, giving a larger late-time acceleration rate (indicated by a higher $H_0$). Therefore, decaying or annihilating dark matter models, such as those studied previously in Refs.~\cite{doroshkevich1989large,oguri2003decaying,Wang:2010ma,cirelli2012gamma,wang2012effects,bjaelde2012origin,wang2013lyman,blackadder2014dark,aoyama2014evolution,enqvist2015,poulin2016,bringmann,pandey2019alleviating,vattis2019late,archidiacono2019constraining,clark2020cmb,haridasu2020late,enqvist2020constraints}, offer a tantalizing hope of resolving 
the $H_0$ and $\sigma_8$ tensions simultaneously.

In this paper, we are specifically interested in the class of models where the energy density in dark matter monotonically converts into dark radiation, with the bulk of the activity happening at low redshift (late time). Our motivation is to investigate whether a model where dark matter converts to dark radiation --- henceforth, a \textbf{DMDR} model --- can satisfy the twin requirements of both being favored by the data and helping alleviate the Hubble and $\sigma_8$ tensions.

In general, interacting dark matter models have the potential to resolve the observations in cosmology that might be otherwise difficult to explain in the standard $\Lambda$CDM model.
Because models with beyond-cold-dark-matter particle content often wash out small-scale structure \cite{tulin2018dark,valli2018dark}, they are well positioned to help alleviate the well-documented challenges observed on small scales (the core/cusp, missing-satellites and too-big-to-fail problems of CDM \cite{smallscalechallenges}).
The Integrated Sachs--Wolfe (ISW) effect has been measured to have an amplitude significantly higher than that predicted in $\Lambda$CDM 
when stacking large voids in the large-scale structure \cite{aiola2015gaussian,kovacs2019more}; the decrease of dark matter would suppress the Weyl potential on large scales, thus enhancing the ISW effect and could thus help to explain this. Finally, cosmic rays from unidentified sources, specifically the galactic positron excess at $\sim$ 300 GeV \cite{aguilar2019towards} and the $\sim$3.5 keV \cite{boyarsky2014unidentified} X-ray line from nearby galaxies, have been hypothesized to be sourced by the decay of dark matter \cite{wang2014cosmological,abazajian2017sterile,farzan2019dark,das2020galactic,ishiwata2020probing} (although they may be inconsistent with some specific dark matter particle models \cite{dessert2020dark,bhargava2020xmm}). All of these lines of inquiry motivate further study of the properties of, and constraints on, the classes of models with DMDR conversion. For example, Wang et al. \cite{wang2014cosmological, Wang:2015fia} investigated a decaying dark matter model that could be mapped into the parameter space of the phenomenological DMDR conversion scenario studied in this paper, and showed that their model can mitigate some of the aforementioned small-scale CDM challenges.

On the theory side, dark matter -- dark radiation conversion is predicted in various physically-motivated scenarios \cite{pospelov2009r,aoyama2014evolution,allahverdi2015dark}.
In particle-dark-matter theories, an unstable dark matter component is predicted in various extensions of the Standard Model. For example, in non-minimal supersymmetric models, the dark sector has a spectrum of particles analogous to particles in the Standard Model, and  heavier particles can decay into the lightest supersymmetric particle \cite{ulrich} which could have properties of dark radiation \cite{higaki2012dark}. More generally, beyond-Standard-Model physics including fifth-force type additional interactions, can naturally accommodate dark matter and dark radiation couplings. Some  have proposed such coupled models as a mechanism to solve the 21 cm absorption anomaly seen by the EDGES experiment \cite{edges,bondarenko2020constraining}. Furthermore, inspiraling and colliding primordial black holes (PBHs) --- dark-matter candidates in their own right \cite{carr2016} --- could transfer energy from dark matter to gravitational waves, which are also a form of dark radiation \cite{raidal2017gravitational,bringmann}. PBHs could also evaporate into beyond-standard-model relativistic species through Hawking radiation \cite{masina2020dark}. 
Various constraints on PBH abundance were extensively studied by the dynamical, lensing, evaporation and accretion footprints of the PBHs \cite{carr2016,laha2019primordial}, 
but several mass windows remain unconstrained, and previously closed windows sometimes re-open when revisited with improved analysis tools \cite{clesse2018seven,montero2019revisiting,smyth2020updated}. 

Any of the aforementioned theoretical models could underlie a phenomenological dark matter-dark radiation conversion model. The key signature of such a model, compared to the standard $\Lambda$CDM model, is the decreased fraction of dark matter in favor of both dark radiation and dark energy. 

Our goal is to study a phenomenological cosmological DMDR model using state-of-the-art cosmological observations. In this work we utilize the CMB temperature, polarization, and lensing potential angular power spectra measured by Planck  \cite{planck2018}, together with type Ia supernovae from Pantheon \cite{jones2018measuring}, baryon acoustic oscillations (BAO) from the BOSS \cite{bossdata}, MGS \cite{mgs}, and 6dFGS \cite{6dfgs} surveys, and tomographic galaxy clustering and weak lensing measured by the Dark Energy Survey (DES) \cite{desy1}. 

This work is presented as follows. We  introduce our DMDR model in section~\ref{model}, stressing its signatures in the CMB and matter power spectrum. In section~\ref{method}, we present the details of our analysis pipeline, including the datasets we use and the theoretical predictions of the DMDR model. In section~\ref{results}, we report combined constraints on the DMDR model from DES-Y1 and external data, along with model comparison between DMDR and $\Lambda$CDM.  We conclude in section~\ref{sec:concl}.

\section{The DMDR model}
\label{model}

\begin{figure}[t]
\includegraphics[width=0.5\textwidth]{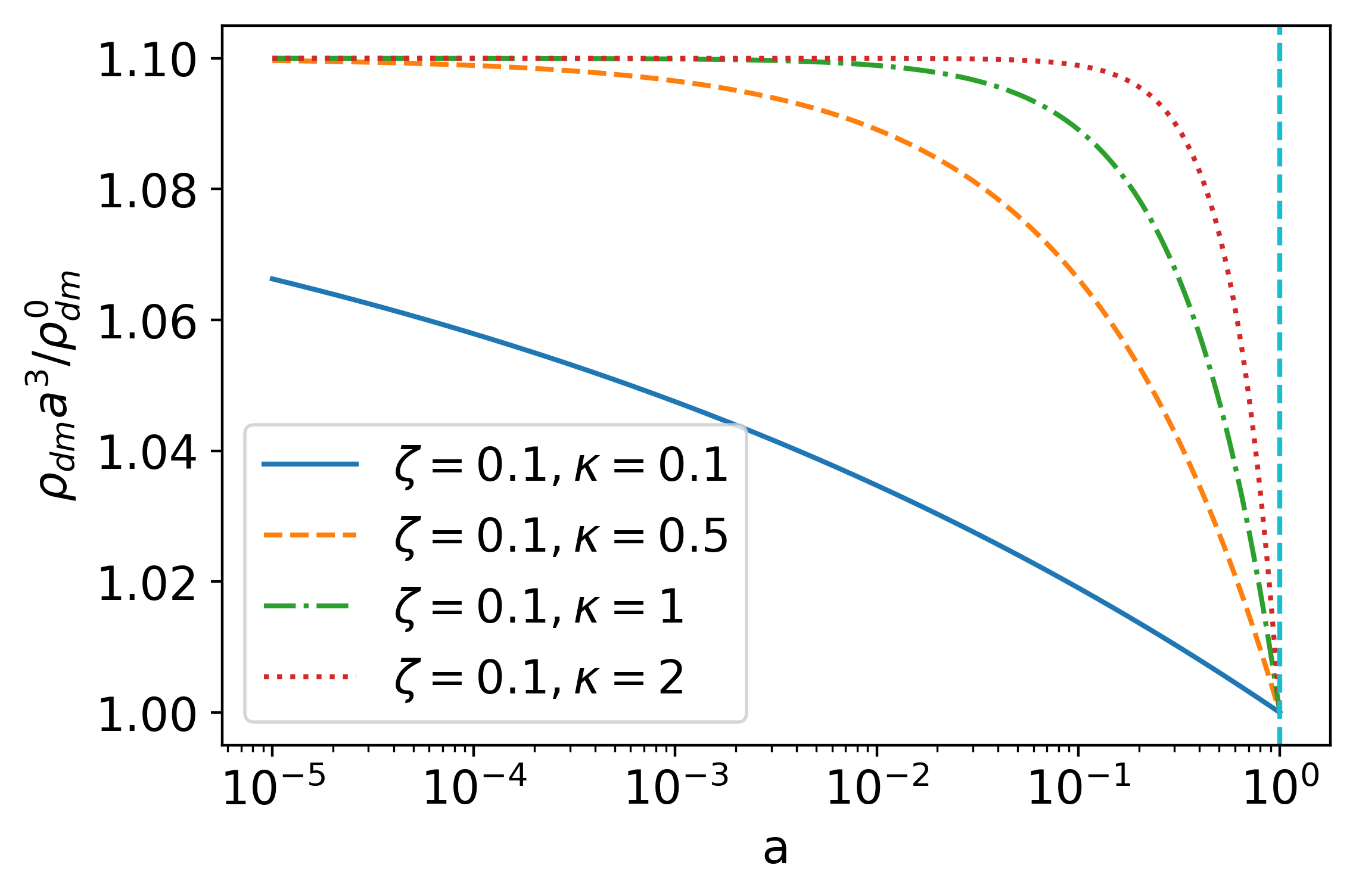}
\caption{Temporal evolution of the comoving dark matter density (in units of  current dark matter density $\rho_{\rm DM}^0$. The legend shows the assumed values of $\zeta$, the fraction of dark matter that has converted into dark radiation since the early universe relative to current density, and $\kappa$, the conversion rate of dark matter. We fixed the standard cosmological parameters to their fiducial values as reported in section~\ref{sec:backgroundeqs}.}
\label{fig:dm}
\end{figure}

Our specific implementation of the dark matter -- dark radiation conversion model is based on the phenomenological model studied by Bringmann et al.\ \cite{bringmann}, hereafter B18. We focus on the case where the conversion process accelerates in time, and the major departures from $\Lambda$CDM happen at late times, as shown in figure \ref{fig:dm}. To obtain a phenomenological model with this behavior, we impose an additional boundary condition onto the original B18 three-parameter ansatz to obtain a steeper rate of dark matter conversion in the recent past ($z\lesssim 10$); see the next subsection. 
Overall, our DMDR model introduces two additional parameters compared to $\Lambda$CDM. 

We now describe the background equations for the model, followed by the description of its perturbations.

\subsection{Background Equations}
\label{sec:backgroundeqs}

The background evolution of the DMDR model is specified by the ansatz of the decreasing dark matter density and the modified continuity equation
\begin{eqnarray}
\rho_{\rm dm}(a) &=& \frac{\rho^0_{\rm dm}}{a^3}\left[ 1+\zeta\frac{1-a^{\kappa}}{1+\zeta a^{\kappa}}\right]
\label{eq:background_1}
\\[0.2cm] 
\frac{1}{a^3}\frac{d}{dt}(a^3 \rho_{\rm dm}) &=& -\frac{1}{a^4}\frac{d}{dt}(a^4 \rho_{\rm dr})=-\mathcal{Q}
\label{eq:background_2}
\end{eqnarray}
where $\rho_{\rm dm}$ and $\rho_{\rm dr}$ are dark matter and dark radiation energy densities, $\rho_{\rm dm}^0$ is the dark matter density today, $a$ is the scale factor, and we introduce
 two new parameters\footnote{ The original ansatz in B18 has three parameters: $\zeta$, $\kappa$, $a_t$, where the last parameter is the characteristic scale factor when the conversion happened. Here we set the mathematical condition $\rho_{\rm dm} a^3=0$  as $a\rightarrow \infty$ to obtain an accelerated decreasing curve near $a=1$. This condition leads to an identity among the three parameters, $1-\zeta a_t^{\kappa}=0$.  We then substitute $a_t = \exp(-\log(\zeta)/\kappa)$ back into the B18 ansatz, arriving at our equation~(\ref{eq:background_1})  which contains the remaining parameters $\zeta$ and $\kappa$. Keeping $\zeta$ or $a_t$ in our model is equivalent; we opted for $\zeta$ based on the fact that it is the more physically intuitive parameter in this case.}:
\begin{enumerate}
    \item $\zeta$, the total amount of dark matter that has already converted into dark radiation, divided by the amount of dark matter at the current time.
    \item $\kappa$, the parameter characterizing the conversion rate. The duration of the conversion roughly corresponds to $O(1/\kappa)$ orders of magnitude change in the scale factor.
\end{enumerate} 
Equation~(\ref{eq:background_1}) provides an ansatz for the time evolution of the comoving density of dark matter. In our late-time DMDR conversion model, the bulk of the conversion occurs around the present time ($a\simeq 1$). 
Equation~(\ref{eq:background_2}) specifies that the energy transfers from dark matter to dark radiation. It also determines the energy transfer flux, $\mathcal{Q}$, as a function of the scale factor $a$, taking the derivative of equation~(\ref{eq:background_1}).

Like the original B18 model, our DMDR model has the generality to cover a wide class of decaying/annihilating dark matter model. For most of the popular decaying/annihilating dark matter models with smooth and simple transition curve, in the $a<1$ region a specific value of $\kappa$ that numerically mimic the transition curve of the dark matter density can be found. Note, since the condition of accelerating conversion rate in the near past is similar to pushing the transition time (labeled by the maximum dark-matter conversion rate) to the future, in the single-body decaying dark matter scenario it suggests a very small decay rate, $\Gamma \ll H_0^{-1}$.

To illustrate the  evolution of background quantities, we first discuss the fiducial cosmological model. We fix the non-DMDR cosmological parameters to the following values based on DES-Y1 fiducial values: matter and baryon densities relative to critical $\Omega_{\rm m}=0.3028$ and $\Omega_{\rm b}=0.04793$, scaled Hubble constant $h=0.6818$, spectral index and amplitude of primordial density fluctuations $n_s=0.9694$ and $A_s=2.198\times 10^{-9}$, physical neutrino density  $\Omega_{\nu}h^2=0.0006155$ (corresponding to the sum of the neutrino masses of $0.058$ eV), and optical depth to reionization $\tau=0.06972$. These parameters, which are common to both DMDR and $\Lambda$CDM models, are also adopted in the illustrations and Fisher forecasts throughout the following sections. We stress that the values of the standard cosmological parameters such as $h$ and $\Omega_{\rm m}$ are by definition set at the present time. Thus the high-$z$ region of the DMDR models in these figures has higher dark matter density. The detailed effect of the DMDR parameters $\zeta$ and $\kappa$ is illustrated in the first batch of Figures in this paper, which we now describe. 

Figure~\ref{fig:dm} shows how the density of dark matter evolves with scale factor, relative to $\Lambda$CDM, for different conversion rates. Varying $\zeta$ scales the curves up and down; in the illustrative plots that follow we choose $\zeta=0.1$. We show the matter density evolution for four different values of the conversion rate $\kappa$; results in figure~\ref{fig:dm} and subsequent figures shows rapid changes in the dark matter density in $a\gtrsim0.1$, suggesting that we may be able to place constraints on such models using current LSS observations.

\begin{figure}
\includegraphics[width=0.5\textwidth]{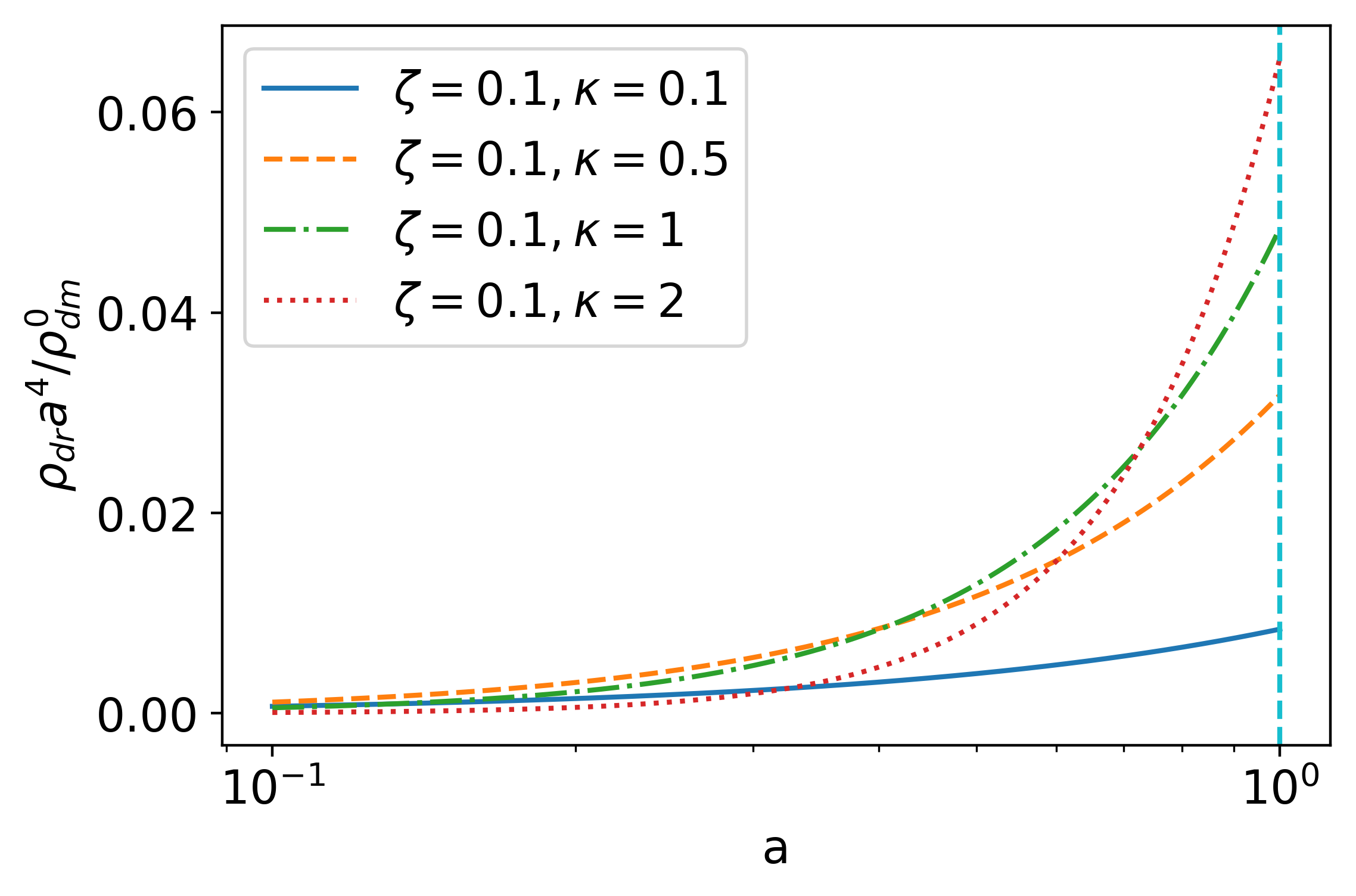}
\caption{Same as figure~\ref{fig:dm}, but now showing the temporal evolution of the dark \textit{radiation} density.}
\label{fig:dr}
\end{figure}

Figure~\ref{fig:dr} shows how the density of dark radiation evolves with scale factor for different conversion rates, relative to $\Lambda$CDM. As the conversion rate parameter $\kappa$ increases, the density of dark radiation in the late universe increases faster. When the dark radiation is produced in the nearer past (for higher $\kappa$), it dilutes less  than if produced over a longer span of time (lower $\kappa$); thus there is more dark radiation at $a=1$ in a larger-$\kappa$ universe. One may worry that large-$\kappa$ models may be automatically ruled out because they apparently lead to a high number of effective relativistic species $\Delta N_{\rm eff} = \rho_{\rm dr}/\rho_{\nu}$, but note that the conversion to dark radiation happens at very low redshifts in our DMDR model and 
thus renders a simple comparison with $\Delta N_{\rm eff}$ constraints derived from the CMB impossible. Hence a detailed analysis of the combination of CMB, LSS and geometric probes is necessary. A more direct impact of dark radiation will be on the expansion history, however, and this will be constrained by the supernova data in our analysis. For the hypergeometric function required to calculate the background density of the dark radiation, we used the special function routine 
from Ref.~\cite{jin1996computation}.

\begin{figure}
\includegraphics[width=0.5\textwidth]{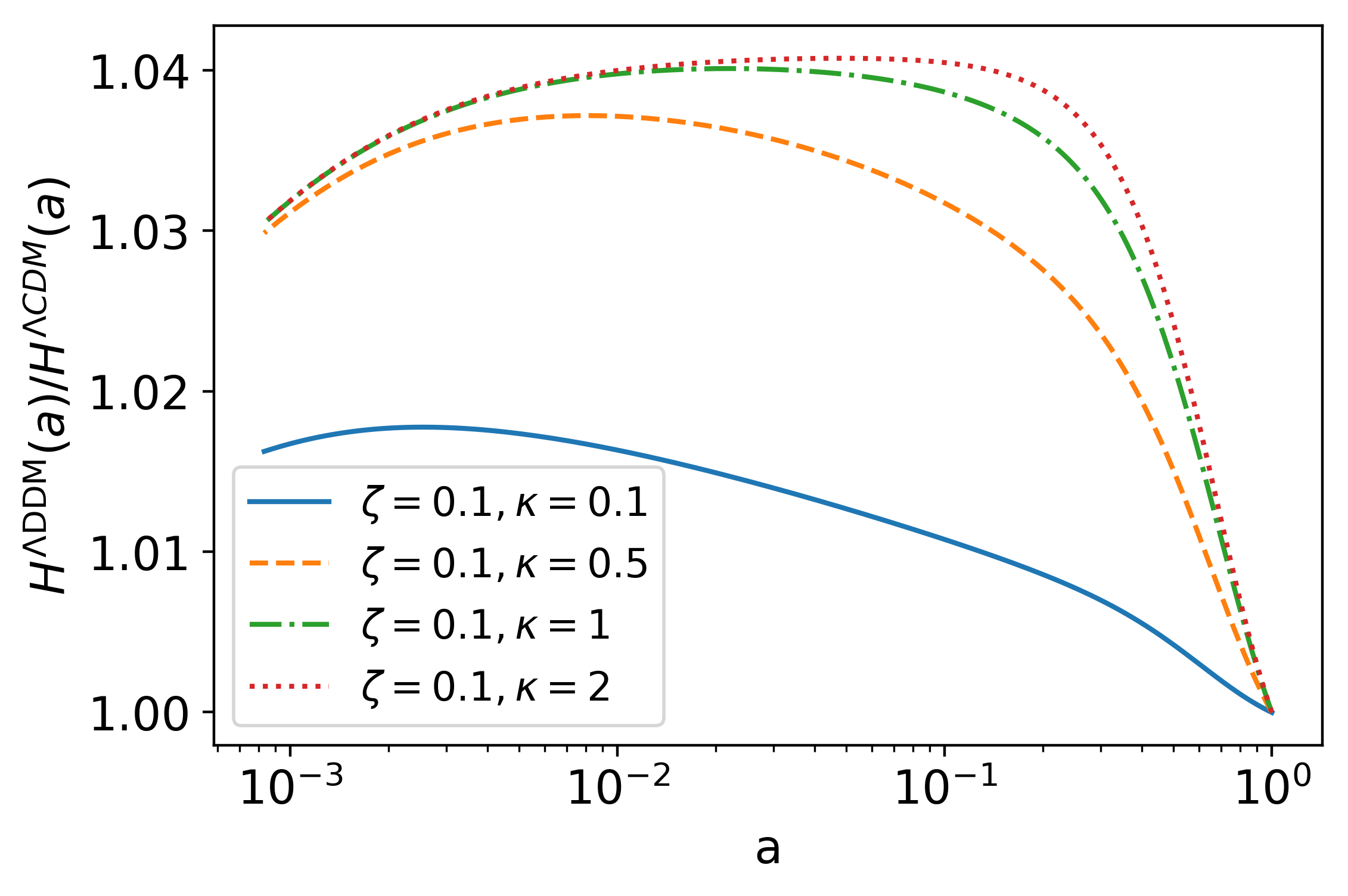}
\caption{Same as figure~\ref{fig:dm}, but now showing time evolution in the ratio between DMDR and $\Lambda$CDM Hubble parameter. }
\label{fig:hubble}
\end{figure}

Figure~\ref{fig:hubble}  shows how the Hubble expansion rate evolves with scale factor for different conversion rates, relative to $\Lambda$CDM. Note that we implicitly hold the present-day values of $\Omega_m$ and $h$ constant in this plot. Then, increasing the conversion rate of dark matter $\kappa$ \textit{increases} the amount of dark matter at $a<1$ relative to today, and hence leads to a more rapid expansion rate, so that $H^{\rm DMDR}(a)/H^{\rm LCDM}(a) > 1$ as seen in figure~\ref{fig:hubble}.



\subsection{Perturbation Equations }

In order to get the matter and radiation perturbation power spectra, we next need to write down the linear perturbation equations of motion for both dark matter and dark radiation, then implement them in the Boltzmann numerical solver CAMB \cite{camb}. We adopt the synchronous gauge throughout this section, following the convention of CAMB. The metric perturbation in synchronous gauge is \cite{maperteq}:
\begin{equation}
    ds^2 = a^2(\tau)[-d\tau^2+(\delta_{ij}+h_{ij})dx^i dx^j],
\end{equation}
where $\tau$ is the comoving time, and $h_{ij}$ with $i,j=1,2,3$ is the metric perturbation.

Most often,  dark radiation is treated as a new species of massless neutrinos (e.g.\ \cite{cambnotes,bringmann}). This conjecture works fine in the scenario with no massive neutrinos, but it produces an incorrect  matter power spectrum that evolves discontinuously away from $\Lambda$CDM when \textit{massive} neutrinos are present. Such behavior is expected because dark radiation (unlike the massless neutrinos) does not interact with massive neutrinos nor does it share the same temperature and entropy with them. In CAMB, the distribution of the energy between neutrino species are specified by a set of time-independent degeneracy numbers, but this is not applicable to the model with energy transfer from dark matter to dark radiation.
\footnote{In the all-massless neutrino case the problem of incorrect time-independent degeneracy numbers could be hidden, because there is no need to partition the energy for the massless species sharing the same equation of motion.} Therefore, as long as the model does not allow for dark matter to massless neutrino conversion, the two species are physically distinct and treating dark radiation as a new type of a massless neutrino is incorrect. Thus we choose to treat dark radiation as an independent perturbation component in the Boltzmann equations. 


In our model, we assume the dark matter to always be cold, meaning that the conversion process to the dark radiation does not provide enough recoil kinetic energy to heat up the dark matter. At the same time, dark radiation in our model does not self-interact or dissipate energy via interactions with dark matter, standard-model particles, or photons after their production, so that dark radiation simply free-streams. As a result, the phase-space perturbation equations for the dark radiation differ  from the massless-neutrino ones only by a collision term. Adopting the perturbation-expansion notation from  \cite{maperteq}, we have
\begin{eqnarray}
    dN &=& f(x^i,P_j,\tau)  dx^1dx^2dx^3 dP_1 dP_2 dP_3   \\[0.2cm]
    f(x^i,P_j,\tau) & = & f_0 (q) \left[ 1+ \Psi(x^i,q,n_j,\tau)\right]\\[0.1cm]
    F(\vec{k},\hat{n},\tau) & = & \frac{\int q^2 dq q f_0(q) \Psi(\vec{k},q,\hat{n},\tau)}{\int q^2 dq q f_0(q)}
\end{eqnarray}
where $x^i$ are comoving coordinates, $P_i$ are their conjugate momentum, $d N$ is the particle number in the phase space differential volume. Here the momentum variable $P_i$ is replaced by $q$ and $n_i$ variables through $P_i = (\delta_{ij}+\frac{1}{2}h_{ij})q n_j$ in the second equation, and $k$-space is Fourier transformed from $x$-space. 

The dark radiation phase-space equation of motion reads
\begin{eqnarray}
\label{eq:dr_eom}
    \frac{\partial F_{\rm dr}(\vec{k},\hat{n},\tau)}{\partial \tau} &+& ik\mu F_{\rm dr}(\vec{k},\hat{n},\tau) = -\frac{2}{3}\dot{h}(\vec{k},\tau)\\[0.2cm]
-\frac{4}{3}(\dot{h}(\vec{k},\tau)&+&6\dot{\eta}(\vec{k},\tau))P_2(\hat{k}\cdot \hat{n}) + \left( \frac{\partial F_{\rm dr}(\vec{k},\hat{n},\tau)}{\partial \tau}\right)_C,
\nonumber
\end{eqnarray}
 where $( \partial F_{\rm dr}(\vec{k},\hat{n},\tau)/\partial \tau)_C$ is the additional collision term due to the conversion between dark matter and dark radiation, to be contrasted with the collisionless massless neutrino equations.
 
We adopt a simple form for the collision perturbation equation involving no dependence on polarization or momentum anisotropy. Specifically,
\begin{equation}
    \left( \frac{\partial F_{\rm dr}(\vec{k},\hat{n},\tau)}{\partial \tau}\right)_C = \frac{\mathcal{Q}(a)a}{\rho_{\rm dr}(a)} (-F_{\rm dr}(\vec{k},\hat{n},\tau)+\delta_{\rm dm}(\vec{k},\tau)).
    \label{eq:collision}
\end{equation}
where $\mathcal{Q}$ is defined in equation~(\ref{eq:background_2}).
When writing down the equation~(\ref{eq:collision}), we adopted the minimal form for the perturbation variation of the conversion term $\mathcal{Q}$:

\begin{equation}
    \delta \mathcal{Q}=\mathcal{Q}\delta_{\rm dm}.
    \label{eq: deltaq}
\end{equation}

In principle, the form of $\delta \mathcal{Q}$ is determined by the microphysics of the dark matter-dark radiation conversion process. The minimal form above has been adopted by previous literature \cite{Ichiki_2004,enqvist2015, poulin2016}, and B18 has demonstrated that the current generation cosmology observations do not have high enough precision to distinguish the detailed $\delta \mathcal{Q}$ perturbation, by carrying out case studies on Sommerfeld enhancement and single-body decay process.

After harmonic expansion of equation~(\ref{eq:dr_eom}),
we get the hierarchy equations for dark radiation. Along with the dark matter perturbation equations, the full set of perturbation equations in DMDR model reads \cite{maperteq,cambnotes,classII,audren2014strongest}: 

\begin{widetext}
\begin{alignat}{2}
\delta'_{\rm dm} + k\mathcal{Z}&=  \frac{a}{\bar{\rho}_{\rm dm}}(\mathcal{Q}\delta_{\rm dm}-\delta \mathcal{Q})=0 & \qquad\qquad [\mbox{Dark Matter}]
\label{eq:Boltzmann}\\
\delta'_{\rm dr} &= -\frac{4}{3}k \mathcal{Z} -k q_{\rm dr} - \frac{a\mathcal{ Q}}{\bar{\rho}_{\rm dr}}(\delta_{\rm dr}-\delta_{\rm dm}) &  [\mbox{Dark Radiation}, \ell=0] \label{eq:hierarchy1}\\
q_{\rm dr}' &=  \frac{k}{3}\delta_{\rm dr}-\frac{2}{3}k\beta_2\pi_{\rm dr} - \frac{a\mathcal{ Q}}{\bar{\rho}_{\rm dr}}q_{\rm dr}  & [\mbox{Dark Radiation}, \ell=1] \label{eq:hierarchy2}\\
\pi_{\rm dr}' &= \frac{2}{5}k q_{\rm dr}-\frac{3}{5}k\beta_3 J^{\rm dr}_3+\frac{8}{15}k\sigma - \frac{a\mathcal{ Q}}{\bar{\rho}_{\rm dr}}\pi_{\rm dr} & [\mbox{Dark Radiation}, \ell=2] \label{eq:hierarchy3}\\
J^{\rm dr'}_{\ell} &=  \frac{k}{2\ell+1}
[\ell J_{\ell-1}^{\rm dr}-\beta_{\ell+1}(\ell+1)J_{\ell+1}^{\rm dr}] - \frac{a\mathcal{ Q}}{\bar{\rho}_{\rm dr}}J_{\ell}^{\rm dr},  & [\mbox{Dark Radiation}, \ell>2]
\label{eq:hierarchy4}
\end{alignat}
\end{widetext}
where $J_{\ell}$ are the harmonic expansions of the phase space perturbation, $J^{\rm dr}_0 \equiv \delta_{\rm dr}$, $J^{\rm dr}_1 \equiv q_{\rm dr} = \frac{4}{3}\theta_{\rm dr}/k$, $J^{\rm dr}_2 \equiv \pi_{\rm dr}=\Pi^{\rm dr}/\bar{\rho}_{\rm dr}$ in CAMB convention;  $\mathcal{Z}$ and $\sigma$ are the metric perturbation coefficients, and $\beta_{\ell}$ are the harmonic expansion coefficients of the gradient operator defined in reference~\cite{cambnotes}. 
Further details of this derivation are included in Appendix~\ref{app:perturbation}. 

The modifications described above are relevant for the continuity equations. For the Einstein equations, the correction is rather straightforward: we simply add the dark-radiation perturbations to the total energy-momentum perturbations.

\subsection{CMB and Matter Power Spectrum}

We now have the ingredients necessary to numerically compute the CMB polarized temperature anisotropies and matter perturbation power spectra, and thus derive the observable quantities that can be compared to data. We implement the background and perturbation equations in the previous two subsections in the Einstein--Boltzmann code CAMB \cite{camb} which is used in the {\fontfamily{qcr}\selectfont
cosmosis} pipeline that we discuss in more detail below.\footnote{DMDR-CAMB using the background and perturbation equations in this work can be found here: https://bitbucket.org/anqich/ddm-camb/src/master/. Please email the corresponding author to get access if it is needed.}  

\begin{figure*}
\includegraphics[width=0.48\textwidth]{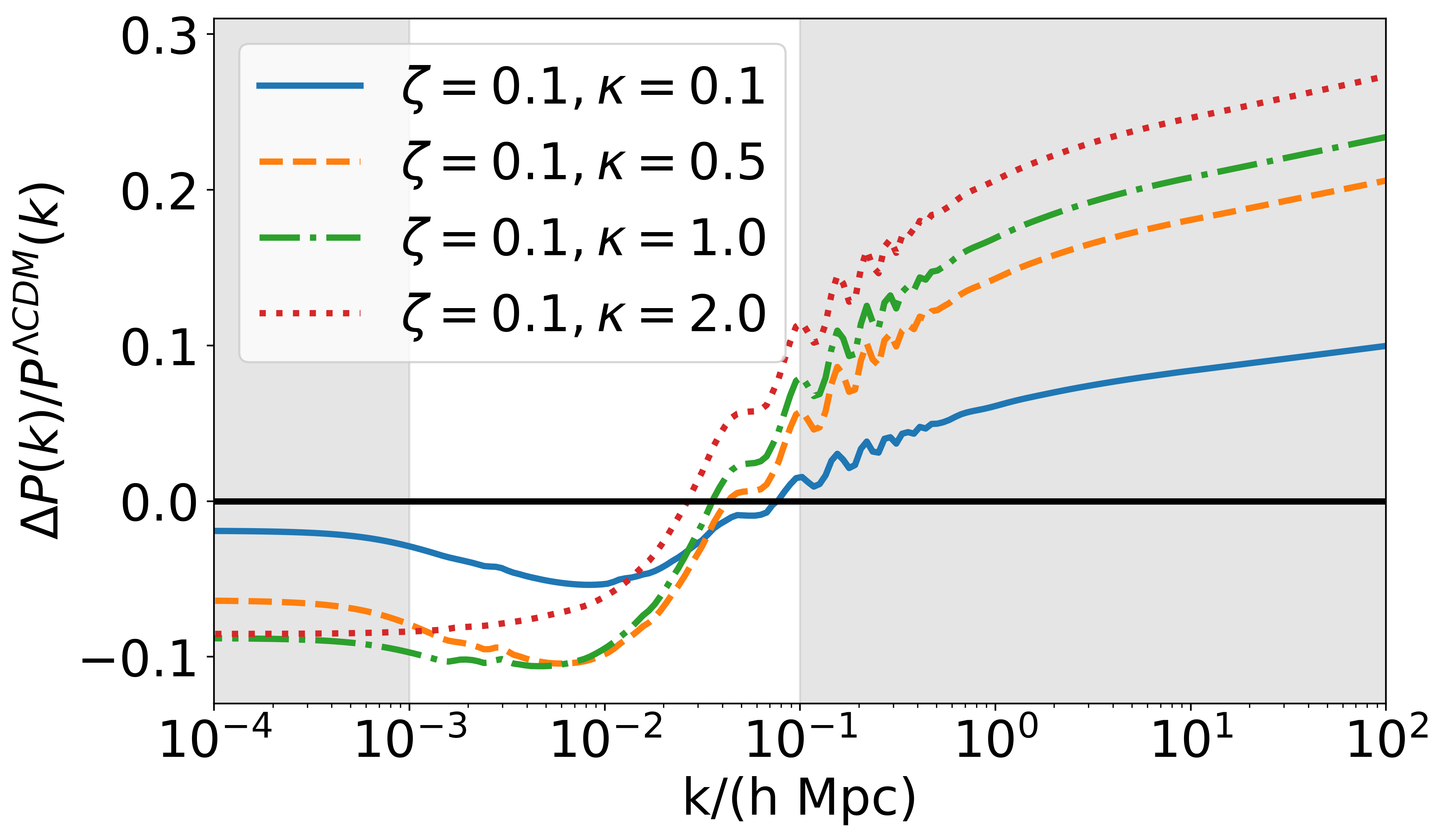}
\includegraphics[width=0.48\textwidth]{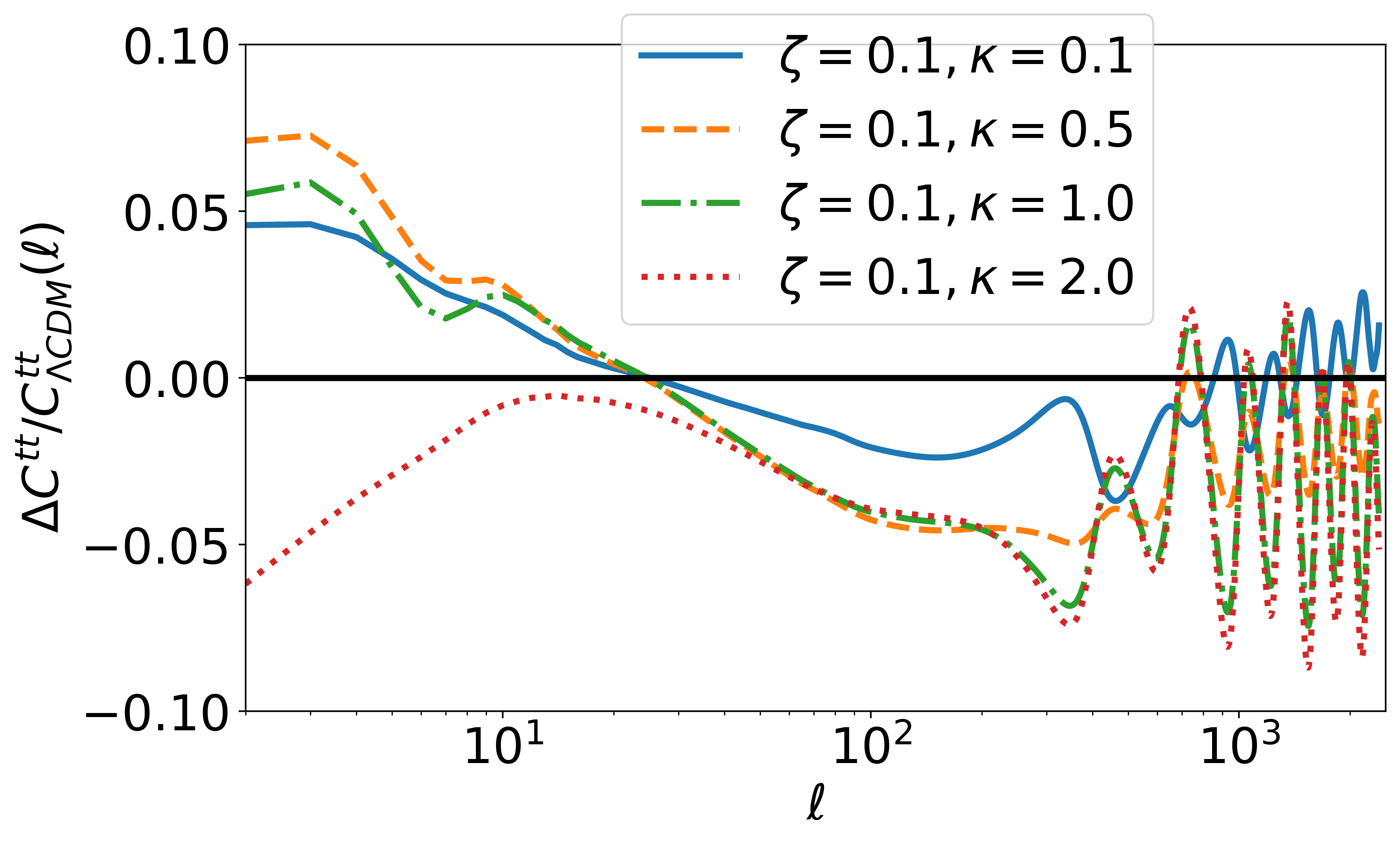}
\caption{Relative difference in the matter power spectrum (left panel) and CMB TT spectrum (right panel) between DMDR and $\Lambda$CDM. We explore the same four sets of ($\zeta, \kappa$) values as in the previous three figures. In the left panel, the white region (between the two shaded regions) denotes roughly the scales used by the DES 3x2pt analysis.
}
\label{fig:pkrel}
\end{figure*}

Figure~\ref{fig:pkrel} illustrates the relative differences between the DMDR and $\Lambda$CDM matter power spectra and their CMB spectra. As with the background-evolution illustrations above, we fix the parameters common to both DMDR and $\Lambda$CDM model to their fiducial values listed in section~\ref{sec:backgroundeqs}, and we only vary DMDR-specific parameters $\zeta$ and $\kappa$. This  ensures that the two cosmologies always converge at late times (see also figure~\ref{fig:dm}). In the early universe, DMDR has more dark matter than $\Lambda$CDM, and this makes the matter and CMB power spectra resemble those in a $\Lambda$CDM cosmology, but with more dark matter. This, in turn, shows up as the small-scale power enhancement, as well as the phase shift in the case of the CMB power spectrum. 

A distinctive feature in DMDR is the dip in the matter power spectrum at $k\sim 10^{-2} \hmpcinv$, the scale corresponding to the horizon crossing at matter--radiation equality. This feature is mostly due to the different expansion history in a higher dark-matter density universe in DMDR. Although we see an increase in the matter power around $k \sim 0.1 h/\rm Mpc$, and might worry that it could boost the amplitude of mass fluctuations $\sigma_8$ and thus exacerbate the LSS tension with CMB, note that this is not the case because we have artificially held most of the cosmological parameters fixed. In fact, DMDR can be qualitatively compared and contrasted with the early dark energy (EDE) models \cite{poulin2019early, hill2020early}. While the EDE models which have a larger dark-matter-to-dark-energy ratio 
after recombination than $\Lambda$CDM, the DMDR model have a \textit{smaller} such ratio relative to $\Lambda$CDM. This works in the direction of reconciling the $\sigma_8$ tension. 

In the CMB temperature power spectrum shown on the right in figure~\ref{fig:pkrel}, the decreasing dark matter density leads to an increase in the late integrated Sachs--Wolfe (ISW) effect caused by the decrease of the gravitational potential as dark matter converts into dark radiation (an exception is the $\kappa = 2$ case which we discuss separately below). Late-ISW effect is caused by the decrease of Weyl potential in the dark-energy-dominant epoch as the expansion of universe accelerates. In $\Lambda$CDM, the decrease of the Weyl potential only happens in the dark-energy-dominated epoch while the potential remains constant in dark matter epoch, but in the DMDR model the late-ISW effect also accumulates in the dark-matter-dominated epoch. This is because the Weyl potential is mainly contributed to by dark matter and a decreasing comoving density of dark matter leads to a decreasing Weyl potential even before dark energy takes over.
Although DMDR imprints in the late-ISW effect are probably buried in the cosmic variance, it does gives these models an additional signature that can be sought in e.g.\ studies of the ISW imprints in the large voids \cite{kovacs2019more}.

The red curve in figure~\ref{fig:pkrel} requires further discussion. This is the case where the dark matter converts at very late times ($z\simeq O(1)$) and rapidly. Therefore, the increased dark-energy-to-dark-matter ratio that is characteristic of DMDR model occurs too late for the late-time ISW to fully benefit from it. In addition, a DMDR model with the same present-day $\Omega_m$ as a $\Lambda$CDM model has more matter relative to dark energy at $z>0$; therefore, contributions to late-time ISW occur later in DMDR than in $\Lambda$CDM. These two effects combine to severely suppress the late-time ISW effect in high-$\kappa$ DMDR models.

\begin{figure}
    \centering
    \includegraphics[width=0.48\textwidth]{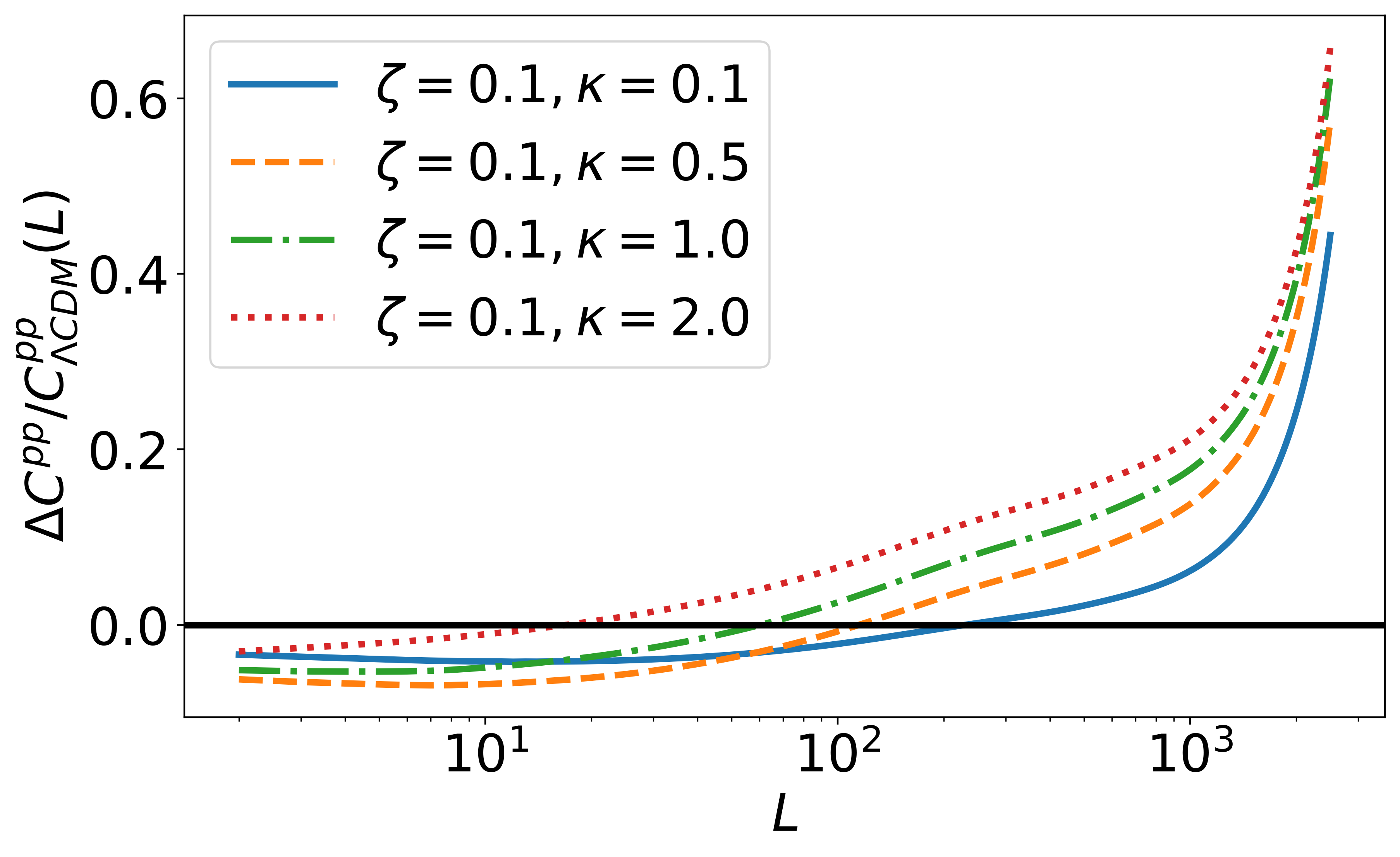}
    \caption{Relative difference in the CMB lensing potential spectrum between DMDR and $\Lambda$CDM, as a function of $\kappa$ for $\zeta = 0.1$. }
    \label{fig:lensing}
\end{figure}

Lastly, we also present the DMDR effect on the lensing potential power spectrum for CMB; see figure~\ref{fig:lensing}. We observe an increase of the lensing potential at small scales (large multipoles $L$) that mimics the amplified large $k$ modes of matter power spectrum seen in figure~\ref{fig:pkrel}. 

\subsection{Nonlinear Matter Power Spectrum Strategies and DES-Y1 Scales Used}
\label{nonlinear}

Obtaining accurate theoretical predictions for nonlinear clustering
in cosmological models outside of $\Lambda$CDM is typically challenging, as these predictions require running suites of  cosmological simulations designed specifically for the extended models. This situation can be contrasted to that in $\Lambda$CDM (and its simplest extension that assume a free but constant dark energy equation of state, wCDM), where the modeling of nonlinear matter power spectrum has been extensively studied with N-body simulations \cite{heitmann2010coyote,owls,mcalpine2016eagle} and analytical fits or models \cite{Smith:2002dz,Bird:2011rb,takahashi,hmcode}.  Limited previous studies of the small-scale structure formation in DMDR include simulations of a less general class of decaying dark matter models than the one we adopt here \cite{enqvist2015}, and the demonstration that relativistic species have negligible contribution to the gravitational physics of the small-scale structure formation \cite{dakin2019}.
One potentially useful alternative to running simulations is recent work \cite{cataneo2019road} which proposes to accurately model beyond-$\Lambda$CDM models by suitably rescaling the $\Lambda$CDM result in order to get one into the desired new model. 
These results are potentially useful and we may study and implement some of them in the future, but they are currently not validated to the level sufficient to enable us to model the nonlinear clustering in our DMDR cosmological model. 

We therefore choose to limit our analysis to purely-linear scales, thus following the same strategy as in the DES-Y1 modified gravity analysis \cite{desy1ext} (see also reference~\cite{Ade:2015rim}). 
To summarize, we start with the difference between the nonlinear
and linear-theory predictions of the observed data in the standard $\Lambda$CDM model at best-fit
values of cosmological parameters,  $\mathbf{d}_{\rm NL}- \mathbf{d}_{\rm lin}$. Using also the full error covariance of DES-Y1, $ \mathbf{C}$, we
calculate the quantity
\begin{equation}
\Delta\chi^2 \equiv (\mathbf{d}_{\rm NL}-\mathbf{d}_{\rm lin})^T\,
\mathbf{C}^{-1}\,(\mathbf{d}_{\rm NL}-\mathbf{d}_{\rm lin})
\end{equation}
and identify the single data point that contributes most to this quantity. We
remove that data point, and repeat the process masking out $\mathbf{d}_{\rm NL}<\mathbf{d}_{\rm lin}$ region until $\Delta\chi^2<1$. 
The resulting set of 334 (compared to the DES-Y1 3x2pt baseline 457) data points that remain
constitutes our fiducial choice of linear-only scales. 

\subsection{Expectations and Forecasts}

Before analyzing the data, we perform a forecast of the expected constraints. We do so in order to understand the parameter degeneracy structure, especially in regards to the new parameters $\zeta$ and $\kappa$. We would also like to understand what constraints are expected on these parameters. However, not all the likelihoods we plan to use in the real-data analysis have the corresponding mock likelihoods available. So for the forecast, we only use the DES-Y1 3x2pt and the Planck 2018 TT-TE-EE-lite data centered at the fiducial $\Lambda$CDM cosmology.
    The likelihood of simulated Planck data vector was calculated by implementing a wrapper of  the work of reference~\cite{planckpy} in {\fontfamily{qcr}\selectfont
cosmosis}.

\begin{figure}
\includegraphics[width=0.5\textwidth]{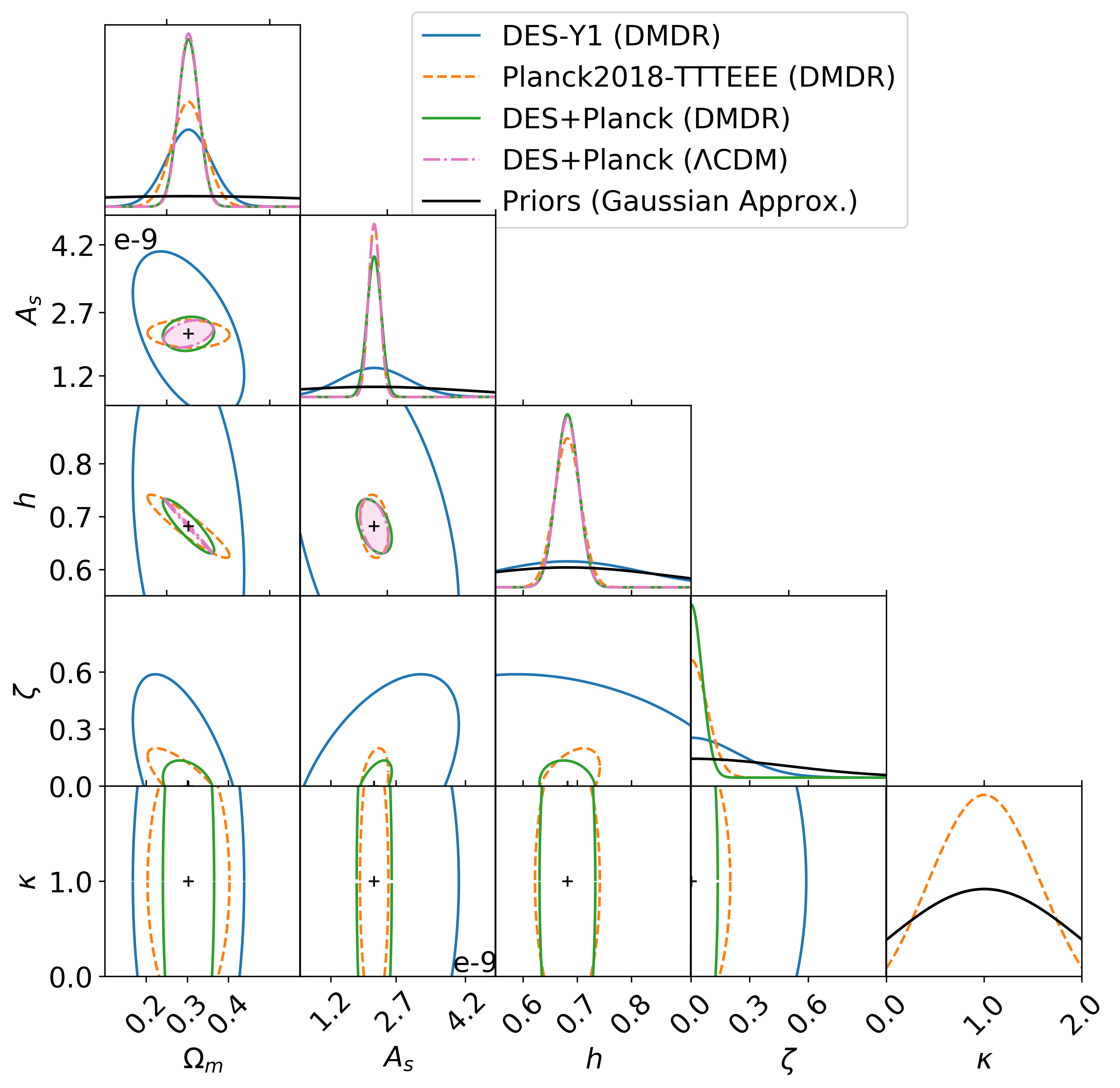}
\caption{The DMDR Fisher forecasts showing 95\% C.L.\ contours assuming simulated DES-Y1 3x2pt data, simulated Planck 2018 data, and the combination of both, all generated close to $\Lambda$CDM cosmology. The forecast is done assuming a Gaussian surface around the fiducial $\Lambda$CDM cosmology, specified by the same parameters in section~\ref{sec:backgroundeqs}. The combined datasets noticeably increased the constraint power, especially on the fraction of converted dark matter $\zeta$. The $\Lambda$CDM model's degeneracy between $h$ and $\Omega_m$ (note a very thin red contour in that plane) opened up in DMDR.} 
\label{fig:ddmfisher}
\end{figure}

To obtain the forecasts on parameter constraints, we adopt the Fisher matrix methodology. The Fisher matrix is defined as
\begin{equation}
    \mathcal{F}_{ij} = \sum_{mn} \frac{\partial v_m}{\partial p_i} [C^{-1}]_{mn} \frac{\partial v_n}{\partial p_j} + [\mathcal{I}^{-1}]_{ij}
\end{equation}
evaluated at the fiducial cosmology, where $v_{m}$ are the theoretically predicted data values, $p_{i}$ are the cosmological and nuisance parameters, $C_{ij}$ is the covariance matrix of the data, and $\mathcal{I}_{ij}$ is the covariance matrix of parameter priors. Fisher matrix calculations typically  incorporate Gaussian priors on the parameters. Because we have flat priors on some of our parameters (see table~\ref{table:pars}), we adopt Gaussian priors of which the \textit{variance} scales with the range (hence variance) of the flat priors that we have. Such Gaussian prior approximations are illustrated by black lines in figure~\ref{fig:ddmfisher}. Thus we add $\mathcal{I}_{ij} = \delta_{ij}{\rm Var}[\mathcal{P}(p_i)]$, where $\delta_{ij}$ is the Kronecker Delta and $\mathcal{P}(p_i)$ any one of the Gaussian approximation of the flat priors from table~\ref{table:pars}. We center the cosmological parameters at the values listed in section~\ref{sec:backgroundeqs}. For the near-fiducial $\Lambda$CDM Fisher calculation, we adopt the DMDR parameter values of $\zeta=10^{-4}$ and $\kappa=1.0$, where all the cosmological observables have negligible difference from $\Lambda$CDM due to small $\zeta$ yet is sensitive enough to the two additional parameters. 
We use the {\fontfamily{qcr}\selectfont
cosmosis}\footnote{https://bitbucket.org/joezuntz/cosmosis/wiki/Home} \cite{cosmosis} Fisher sampler to forecast the constraints on the DMDR parameters. 

In the Fisher forecast results shown in figure~\ref{fig:ddmfisher}, we observe that:
\begin{itemize}
    \item The DMDR model breaks the tight correlation between $\Omega_{\rm m}$ and $h$ for {\it Planck}. In $\Lambda$CDM $\Omega_{\rm m}$ and $h$ are strongly anticorrelated because $\Omega_{\rm m} h^2$ is tightly constrained by the morphology of the acoustic peaks in the CMB spectrum. In DMDR,  the background evolution has more freedom given by the variation of $\zeta$ and $\kappa$, thus weakening this degeneracy by adding more degrees of freedom in this 2D space. 
    \item Furthermore, DES has a different degeneracy direction from Planck in the $\Omega_{\rm m}$--$h$ plane, so that when the two probes are combined the  degeneracy in this space is significantly reduced. Because $\zeta$ is significantly correlated with $\Omega_{\rm m}$, this degeneracy breaking greatly helps in constraining $\zeta$. 
    \item In figure \ref{fig:ddmfisher} we assumed a DMDR cosmology very close to $\Lambda$CDM (with $\zeta = 10^{-4}$). In that case, there is effectively no constraint on the conversion rate $\kappa$, as expected. 
\end{itemize}
Note again that the Fisher forecasts above are centered at $\zeta = 10^{-4}$, $\kappa = 1.0$ (near) $\Lambda$CDM. We have checked that, as the fiducial values of both $\zeta$ and $\kappa$ increase away from their $\Lambda$CDM values of zero, the forecasted constraints strengthen. Such behavior in Fisher matrix forecasts is not uncommon and occurs when the dependence of the measured quantities on the parameters of interest is nonlinear. Nevertheless, the constraints presented in figure~\ref{fig:ddmfisher} give us a rough idea of what to expect from the real data. We have also checked that increasing the fiducial converted fraction to $\zeta=0.1$ only modestly strengthens constraints on $\kappa$.

We now proceed to describe our data and methodology.

\section{Methodology}
\label{method}
We follow the general scheme for the $\Lambda$CDM extension model analysis of the DES-Y1 3x2pt combined probes, which was described in detail in the DES-Y1 extensions paper~\cite{desy1ext}. In this section we will mainly focus on the methodology and systematics tests results specifically for the DMDR model, for full details, see references~\cite{desy1,desy1ext}. 

\subsection{Theory Prediction Pipeline}\label{sec:theory_pipeline}

Our theory predictions for the DES 3x2pt data vector are derived from the 2D projection of the 3D matter and Weyl potential power spectra, incorporating complexities like nonlinear physics, galaxy bias, intrinsic alignments, photo-$z$ bias, and shear calibration bias. The detailed derivation of 3x2pt theory prediction were described in Sec.~IV.A of Ref.~\cite{desy1}. Here we only go through the procedures that are specifically modified for the DMDR model.

We first modify the Boltzmann code CAMB by implementing the equations described in section \ref{model}, and refer to this modified version as DMDR-CAMB. We also add a flag on $\sigma_8$ to ensure numerical stability in the nonlinear subroutine of DMDR-CAMB by attributing zero likelihood to  models with $\sigma_8>1.4$ or $\sigma_8<0.4$. The resulting filter prior  $\sigma_8 \in [0.4,1.4]$, is about  $\sim 10 \sigma$ wide on each side of the fiducial value (relative to the DES-Y1 $\Lambda$CDM analysis \cite{desy1}, $\sigma_8=0.807 ^{+0.062}_{-0.041}$), and thus not expected to affect the overall constraints. 

Next, the relation between the different cosmological quantities in the flat universe is enforced differently in DMDR comparing to $\Lambda$CDM because of a larger fraction of radiation density. 
The flat-universe relation is
\begin{equation}
    \Omega_{\rm m} + \Omega_{\Lambda}+\Omega_{\rm dr} = 1.
\end{equation}
Specifically, while in  $\Lambda$CDM the  flatness condition implies $\Omega_{\Lambda} = 1-\Omega_{\rm m}$,  in flat DMDR we enforce $\Omega_{\Lambda} = 1-\Omega_{\rm m}-\Omega_{\rm dr}$ instead.


Finally we improve upon the usual assumption that the Weyl potential $\Phi$ is completely contributed by matter in the late universe, $\Phi = \frac{3}{2}\Omega_{\rm m} H_0^2\delta_{\rm m}/ac^2$. Recall the Weyl potential defined via the metric potentials $\phi$ and $\psi$ in Newtonian gauge:
\begin{equation}
    \begin{aligned}
    \Phi &= (\phi+\psi)/2\\  
    ds^2 &= a^2(-(1+2\psi)dt^2+(1-2\phi)dx^2).
\end{aligned}
\label{eq:Weyl}
\end{equation}
The assumption that  the $\Phi$ power spectrum is proportional to the matter power spectrum is only reliable for negligible amounts of relativistic species in the late universe, which holds in $\Lambda$CDM but can break in DMDR models with large $\zeta$. At super-horizon scales, $\Phi$ diverges from the local matter perturbation. Our strategy is to take the appropriate ratio between the linear Weyl potential power spectrum $P_{\Phi \Phi}^{\rm lin}$ and the linear matter power spectrum $P_{\delta \delta}^{\rm lin}$, and then modify the shear clustering, galaxy clustering, and galaxy--galaxy power spectra. The Weyl-corrected (WC) power spectra  are:
\begin{eqnarray}
P_{X X}^{\rm WC} &=& R_{\rm Weyl} \, P_{X X}\\[0.2cm]
P_{g X}^{\rm WC} &=& R_{\rm Weyl}^{1/2} P_{g X},
\end{eqnarray}
with the dimensionless Weyl-correction factor defined as
\begin{equation}
    R_{\rm Weyl} \equiv \frac{P_{\Phi \Phi}^{\rm lin}}{\left[\frac{3}{2}\Omega_{\rm m} H_0^2(z+1)^2/c^2\right]^2}\frac{1}{P_{\delta \delta}^{\rm lin}} \label{eq:ratio} 
\end{equation}
where $X\in \{\gamma, {\rm IA}\}$ is a component of the correlation function that needs the Weyl correction (specifically, the shear and intrinsic alignments), and $g$ stands for the galaxy position. Hence $P^{\rm WC}_{X X}, P^{\rm WC}_{g X}$ are building blocks for the corresponding projected (two-dimensional) angular correlation functions; for example $P^{\rm WC}_{\delta\delta}$ is used for the calculation of 2D lensing shear power. The physical reason that the IA and shear components require the gravitational potential correction is that these processes are directly determined by the gravitational field; galaxy shear is formed by the bending of light in the gravitational field, and IA is induced by the tidal gravitational field generated by nearby mass. 

The Weyl potential and Newtonian potential in principle differ because they depend on different gravitational fields. In practice, we find that their relative difference is $<1\%$ throughout the expansion history in a not-strongly-anisotropic metric in both DMDR and $\Lambda$CDM. We are thus justified in calculating the  correction ratio in equation~(\ref{eq:ratio}) from the Weyl-potential power spectrum. 
We further assume that Weyl potential correction is linear and commutes with intrinsic alignments and galaxy bias (this dramatically simplifies the implementation in the code). While this is not guaranteed to be true, given the current linear modeling of intrinsic alignments and galaxy bias any leading-order adjustment is likely absorbed by the nuisance parameters.  Any scale-dependent caveats of this assumption should be further suppressed by the fact that we adopt conservative scale cuts to limit the impact of uncertainties in the modeling of nonlinearities, 

Lastly, as discussed in section~\ref{nonlinear}, we adopt Takahashi et al.\ halofit prescription \cite{takahashi} to produce the nonlinear matter power spectrum. We ensure the robustness of our analysis to small-scale physics by cutting out the data points at nonlinear scales as described in \cite{desy1ext}. 

In Appendix \ref{sec:pipelinecomparison} we include a comparison between Y1 analysis pipeline and our DMDR pipeline when both are applied to the $\Lambda$CDM mock data vector. It illustrates that the pipeline modifications do not induce noticeable bias ($\lesssim 0.1 \sigma$).

\subsection{Parameters and Priors}
\label{prior}
The DES 3x2pt data analysis applied to the DMDR model includes a total of 28 parameters; they are listed in table~\ref{table:pars}. There are eight cosmological parameters and 20 nuisance parameters. DMDR introduces two additional cosmological parameters to the usual six ($\Omega_{\rm m}, h, \Omega_{\rm b}, n_{\rm s}, A_{\rm s}, \Omega_\nu h^2$): the fraction of the converted dark matter  $\zeta$ 
and the dark matter conversion rate $\kappa$. When combining DES 3x2pt data set with the external data sets, three more parameters, the reionization optical depth $\tau$, supernova absolute magnitude $M$, and the Planck-lite likelihood nuisance parameter $a_{\rm Planck}$ are added into the variables. Their priors are presented in table \ref{table:pars_extra}.

The prior on $\zeta$ is flat in the range $\zeta \in [0.0, 1.0]$. This range is bounded by the limit when there is no dark matter conversion, and the limit when half of the dark matter has converted since the primordial time. The latter choice is based on the fact that the early-time Planck measurement of the matter density, $\Omega_{\rm m} = 0.3166 \pm 0.0084$ \cite{planck2018}, is within $20\%$ of the late-time DES measurement, $\Omega_{\rm m} = 0.264^{+0.032}_{-0.019}$. Hence, there is no indication that a large fraction of the dark matter has converted at $z \lesssim 1000$; this conclusion is also in line with previous work \cite{enqvist2015,poulin2016,kumar2018cosmological,clark2020cmb}.

The prior on the conversion rate $\kappa$ is also flat, with the range $\kappa \in [10^{-7},2]$. We set the lower bound very slightly above zero in order to ensure numerical stability of the modified code, and checked that in this small-$\kappa$ limit the observables agree with those of $\Lambda$CDM. The upper prior limit is determined by the fact that neither the matter power spectrum nor the CMB angular power spectrum varies at a detectable level  when  $\kappa >2$. This, in turn, can be understood from the evolution of the dark matter density illustrated in figure~\ref{fig:dm}. When the conversion rate is as high as 2, new physics happened well after recombination and in the late stages of structure formation, allowing the DMDR model to mimic a $\Lambda$CDM universe with a higher density of dark matter. Thus models with $\kappa \gtrsim 2$ display a strong degeneracy between the new parameters $(\zeta, \kappa)$ and $\Omega_{\rm m}$, and are difficult to constrain tightly.  It is important to keep this in mind when interpreting the $\kappa$ posterior when it is pushed to the upper prior bound. 

The cosmological parameters have flat priors that are nearly 
the same as in DES-Y1 (there are a few very minor differences between the two), and the nuisance parameters that model tomographic intrinsic alignments effect, photo-z uncertainty, shear calibration, and galaxy bias  have the same Gaussian priors as in the DES-Y1 3x2 analysis \cite{desy1}. We also impose a hard filter on the derived parameter $\sigma_8$ within $[0.4,1.4]$ as described in section~\ref{sec:theory_pipeline}.  

\begin{table}[t]
\centering
\caption{Cosmological and nuisance parameters in DES-Y1 3x2pt analysis, and their priors.}
\begin{tabular}[t]{lccc}
\hline
Parameter & Prior\\
\hline
\multicolumn{2}{c}{\textbf{Cosmological}}\\
$\Omega_{\rm m}$& flat (0.1, 0.9)\\
$h$& flat (0.55, 0.91)\\
$\Omega_{\rm b}$& flat (0.03,0.07)\\
$n_{\rm s}$& flat (0.87, 1.07)\\
$A_{\rm s}$& flat ($5\times 10^{-10}$, $5\times 10^{-9}$)\\
$\Omega_{\nu}h_0^2$& flat (0.0006, 0.01)\\
$\zeta$& flat (0.0, 1.0)\\
$\kappa$& flat ($1\times 10^{-7}$, 2.0)\\
$\sigma_8\,$ \mbox{(derived)} & $\in$ (0.4, 1.4)\\
\hline
\multicolumn{2}{c}{\textbf{Lens Galaxy Bias}}\\
$b_i, (i=1,...5)$ & flat(0.8, 3.0)\\
\hline
\multicolumn{2}{c}{\textbf{Intrinsic Alignment}}\\
\multicolumn{2}{c}{$A_{IA}(z) = A_{IA}[(1 + z)/1.62]\eta_{IA}$}\\
$A_{IA}$ & flat (-5, 5)\\
$\eta_{IA}$ & flat (-5, 5)\\
\hline
\multicolumn{2}{c}{\textbf{Lens photo-z shift (red sequence)}}\\
$\Delta z_l^1$ & Gauss (0.008, 0.007)\\
$\Delta z_l^2$ & Gauss (-0.005, 0.007)\\
$\Delta z_l^3$ & Gauss (0.006, 0.006)\\
$\Delta z_l^4$ & Gauss (0.00, 0.01)\\
$\Delta z_l^5$ & Gauss (0.00, 0.01)\\
\hline
\multicolumn{2}{c}{\textbf{Source photo-z shift}}\\
$\Delta z_s^1$ & Gauss (-0.001, 0.016)\\
$\Delta z_s^2$ & Gauss (-0.019, 0.013)\\
$\Delta z_s^3$ & Gauss (0.009, 0.011)\\
$\Delta z_s^4$ & Gauss (-0.018, 0.022)\\
\hline
\multicolumn{2}{c}{\textbf{Shear calibration}}\\
$m^i, (i=1,...4)$ & Gauss (0.012, 0.023)\\
\hline
\end{tabular}
\label{table:pars}
\end{table}%

\begin{table}[ht]
\centering
\caption{Additional parameters used in the analysis with external datasets, along with their priors.}
\begin{tabular}[t]{lccc}
\hline
Parameter & Prior\\
\hline
\multicolumn{2}{c}{\textbf{Cosmological}}\\
$\tau$& flat (0.01, 0.2)\\
\hline
\multicolumn{2}{c}{\textbf{Supernovae Parameter}}\\
$M$ & flat ($-20.0$, $-18.0$)\\
\hline
\multicolumn{2}{c}{\textbf{Planck-lite Nuisance Parameter}}\\
$a_{\rm Planck}$ & Gauss (1.0, 0.0025)\\
\hline
\end{tabular}
\label{table:pars_extra}
\end{table}%

\subsection{Datsets}
\label{data}

Our cosmological parameters analysis will be performed on DES-Y1 3x2pt datasets, external datasets, and the combination of all datasets separately.

We first describe the DES-Y1 "3x2pt" measurements; here 3x2pt  refers to three sets of two-point correlation functions as follows. Let $i$ and $j$ denote source-redshift bins (out of four total), and $a$ and $b$ denote the lens bins (out of five total). The correlation functions that form a set of observables that we call the "data vector" are:
\begin{itemize}
    \item $\xi^{ij}_{\pm}(\theta)$, the correlation between galaxy shear measured in source bins $i$ and $j$;
    \item $\gamma^{ib}_t(\theta)$, the cross correlation between the  galaxy shear in source bin $i$ and the galaxy positions in lens bin $a$;
    \item $w^{ab}(\theta)$ the  correlation between galaxy positions in lens bins $a$ and $b$.
\end{itemize}
The five redshift bins of the lens galaxy catalog are processed using redMaGiC \cite{redmagic}
\begin{equation*}
\begin{split}
z = [(0.15 \sim 0.3), (0.3 \sim 0.45), (0.45 \sim 0.6), \\
(0.6 \sim 0.75), (0.75 \sim 0.9)],
\end{split}
\end{equation*}
while the four redshift bins of the source galaxy catalog, obtained
using the process called
\texttt{METACALIBRATION} \cite{metacalibration}, are
\begin{equation*}
z = [(0.2 \sim 0.43), (0.43 \sim 0.63), (0.63 \sim 0.9), (0.9 \sim 1.3)].
\end{equation*}
Each tomographic two-point correlation function has 20 log-spaced angular bins in the range $2.5' < \theta < 250'$, and a total of 45 tomographic angular correlation functions in each theta-bin, for a total of $20 \times 45 = 900$ data points. Cutting out small angular scales to avoid uncertainties with modeling nonlinearities (see section~\ref{nonlinear}) leaves 334 measurements. We refer the reader for other details, including those of theoretical modeling, to \cite{desy1}. Treatment of some details specific for the DMDR is discussed in section \ref{sec:theory_pipeline}.

Now we describe the external datasets that we adopt; they are:
\begin{itemize}
    \item Cosmic microwave background (CMB): {\it Planck} 2018 high-$\ell$ TT, TE, EE, polarization modes temperature spectra with $\ell \geq 30$ from Plik-lite likelihood, and TT, EE of the low-$\ell$, $\ell \leq 29$ from Commander and SimAll likelihood, plus lensing potential $C_{\ell}$s with multipoles $ 8 \leq L \leq 400$ from SMICA likelihood. \cite{planck2018, aghanim2019planck}
    \item Type Ia supernovae: we adopt the binned Pantheon SNe Ia dataset \cite{jones2018measuring} covering the redshift range $0.01<z<2.3$.
    \item Baryon acoustic oscillation (BAO): we adopt the BOSS DR12 \cite{bossdata} measurements of $H r_s/r_s^{\rm fid},D_m r_s^{\rm fid}/r_s$ at redshifts [0.38, 0.51, 0.61], the SDSS-MGS \cite{mgs} measurement of $\alpha = (D_V/D_V^{\rm fid})(r_s^{\rm fid}/r_s)$ at redshift 0.15, and the 6dFGS \cite{6dfgs} measurement of $r_s/D_V$ at redshift 0.106. The BOSS DR12 data come with a full covariance matrix, while all other data points only have diagonal uncertainties.
\end{itemize}

We do not include the redshift space distortion (RSD) measurements that we previously used in the DES+External data analysis \cite{desy1ext}. We make this choice because DMDR allows for a scale-dependent growth of linear density perturbations, and the bias on $f\sigma_8$ measurements could be significant when the default $\Lambda$CDM templates are used in the compression of RSD information in the presence of a scale-dependent growth \cite{taruya2014beyond, barreira2016validating}. 

\subsection{Samplers}

For our principal results --- constraints in the multi-dimensional parameter space --- we use {\fontfamily{qcr}\selectfont
Polychord} \cite{handley2015polychord}.
{\fontfamily{qcr}\selectfont
Polychord} is a nested sampler with outstanding performance on Bayesian evidence estimation, which is useful for tension and model comparison analysis. We set {\fontfamily{qcr}\selectfont
Polychord} live\_points = 250, num\_repeats = 60, and tolerance = 0.1.  This combination of settings was optimized to obtain precise and accurate results --- especially in regards to the Bayesian-evidence computation --- given our available CPU time.

We also need to run a number of chains for our systematic tests (shown further below in figure~\ref{systematics}). High-quality nested-sampler runs are too time-consuming to be used for these runs. We thus make use of a couple of alternative numerical tools. First, we use the {\fontfamily{qcr}\selectfont
Multinest} \cite{multinest} sampler, which is faster than {\fontfamily{qcr}\selectfont
Polychord}. We use the {\fontfamily{qcr}\selectfont
Multinest} sampler with settings live points = 250, efficiency = 0.3 and  tolerance = 0.01. Second, we adopt our own importance sampler.  

We use these two in conjunction as follows. We first run a baseline chain on uncontaminated theory predicted data vector, and save 334 3x2pt data points for each sample in the chain file. For the importance sampling, we re-weight the samples by a factor $w_{\rm new} = [\mathcal{L}_{\rm new}/\mathcal{L}_{\rm old}] w_{\rm old}$, where $\mathcal{L}_{\rm old}$ is the old likelihood from the MCMC chain, and $\mathcal{L}_{\rm new}$ is the new likelihood calculated using the systematics contaminated data vector and the theory 3x2pt saved for the MCMC samples. In this way, the importance sampler can produce a chain for certain systematic tests in minutes, as opposed to days which running the theoretical pipeline at each sample would take. This process is therefore very CPU-time-efficient, but is only valid in cases when importance sampling is representative on the baseline samples, and when the parameter space remains the same. 
Because sample systematics considered in our tests happen to lead to small deviations from the fiducial model 
--- thanks to our adoption of linear-only scales and nuisance parameters to model general systematics --- this assumption is justified.   Quantitatively, the criterion for the effectiveness of the importance sampling is given by the effective sample size (ESS) given by ${\rm ESS} = (\sum w)^2/\sum (w^2)$. We regard importance sampling as trustworthy if post-importance sampling ESS preserves $\gtrapprox 0.8$ of the baseline ESS, and this is satisfied for all of our systematic tests that use importance sampling. 

In summary, for the real data chains we used {\fontfamily{qcr}\selectfont
Polychord} as the sampler. The systematic tests using the importance sampler are the baryonic, non-Limber, magnification and RSD non-Limber effects. The IA systematics are modeled by nuisance parameters, so they cannot use importance sampling. We run {\fontfamily{qcr}\selectfont
multinest} chain for the two IA systematics validation.

Now we proceed to the validation of pipeline robustness against systematics.

\subsection{Systematics Tests}

Systematic errors, both theoretical and observational, are always a worry for large-scale structure analyses. To address this, we adopt a two-pronged strategy. First, we restrict ourselves to linear scales only, as described in section~\ref{nonlinear}. Second, we perform a battery of validation tests by adding various systematic effects to the data and monitoring how the results on the key cosmological parameters change. We now describe this latter strategy.

We start from a noiseless $\Lambda$CDM mock data vector for DES and Planck; that is, corresponding power spectra that contain no stochastic noise and are centered on the concordance theory model. The Planck mock likelihood is 
based on the compressed likelihood work \cite{planckpy}, centered at $\Lambda$CDM fiducial cosmology. The DES likelihood is identical to the one adopted in this analysis, using theory predicted mock data files. We calculate the cosmological constraints from this baseline case. We then add
the systematic effects described in Sec. IV.A of DES-Y1 extended-models paper \cite{desy1ext},
corresponding to baryonic effects, Limber approximation, magnification bias, Limber approximation with redshift space distortion, two intrinsic alignment models and nonlinear galaxy bias, to generate systematics contaminated data vectors. In each of those cases, we redo the cosmological analysis and evaluate the errors on the key parameters. 

\begin{figure}[ht]
\includegraphics[width=0.48\textwidth]{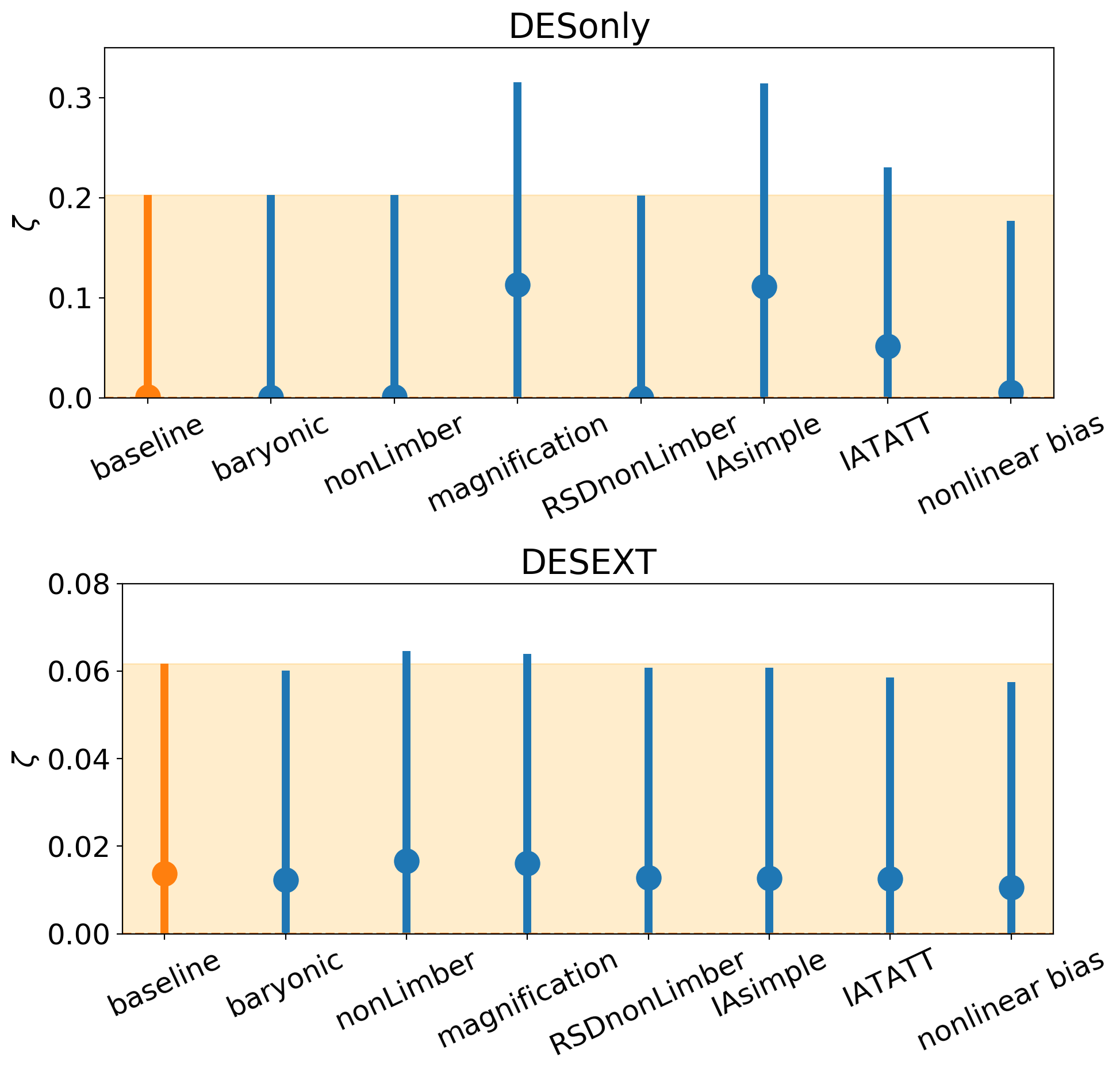}
\caption{The effect of different systematics biases on $\zeta$ for DES-only (top) and DES+EXT (bottom) analysis. The only systematics that show a visible impact are the magnification and intrinsic alignments for the DES-only data, causing a $\approx 0.5 \sigma$  bias on $\zeta$. All other systematics studied here lead to negligible biases. }
\label{systematics}
\end{figure}

The results are shown in figure~\ref{systematics} for the DES-only case (upper panel) and DES+EXTERNAL dataset (lower panel). We see that the systematics are causing at most $0.5 \sigma$ bias in dark matter converted fraction $\zeta$ in DES-only analysis, and no noticeable bias is observed when for the combination of DES and External dataset. The slight deviation ($\sim 0.2 \sigma$) between the best-fit value of $\zeta$ and the assumed $\Lambda$CDM input $\zeta=0.0$ is most likely due to the fact that we ran this test with synthetic DES likelihood but real BAO and supernovae data, and the latter two are not enforced to recover the input-model parameter values.

Because the fiducial simulated data vector is at the $\Lambda$CDM cosmology, $\kappa$ is not constrained and no interesting conclusion could be made on systematic bias. We therefore conclude that our results are robust to some of the key systematic errors, at least to the extent that our systematic models represent the real-world errors.

\subsection{Blinding}
We blinded our real data analysis in the following way. After obtaining the MCMC chain on the real data, before unblinding the cosmological results, we added a random number scaled by the variance of the parameter to the MCMC samples. During the blinded stage of the analysis, we carried out the postprocesses including 2-D contour plots and marginalized parameter constraints on these shifted samples. Our blinding preserves the shape of the contours with random shifting. Thus before proceeding to unblinding, we checked that the contour shapes are reasonable for the data constraining power, and the last few samples have the likelihoods at correct order of magnitude (they are usually not the MAP). In the end we unblinded the cosmological results by resuming the raw samples of the real data MCMC chain. No change to the pipeline was done after unblinding, for the results reported in the next section. The real data analysis pipeline is completely consistent with the systematics test in the above subsection.

\begin{figure*}
    \centering
    \includegraphics[width=0.85\textwidth]{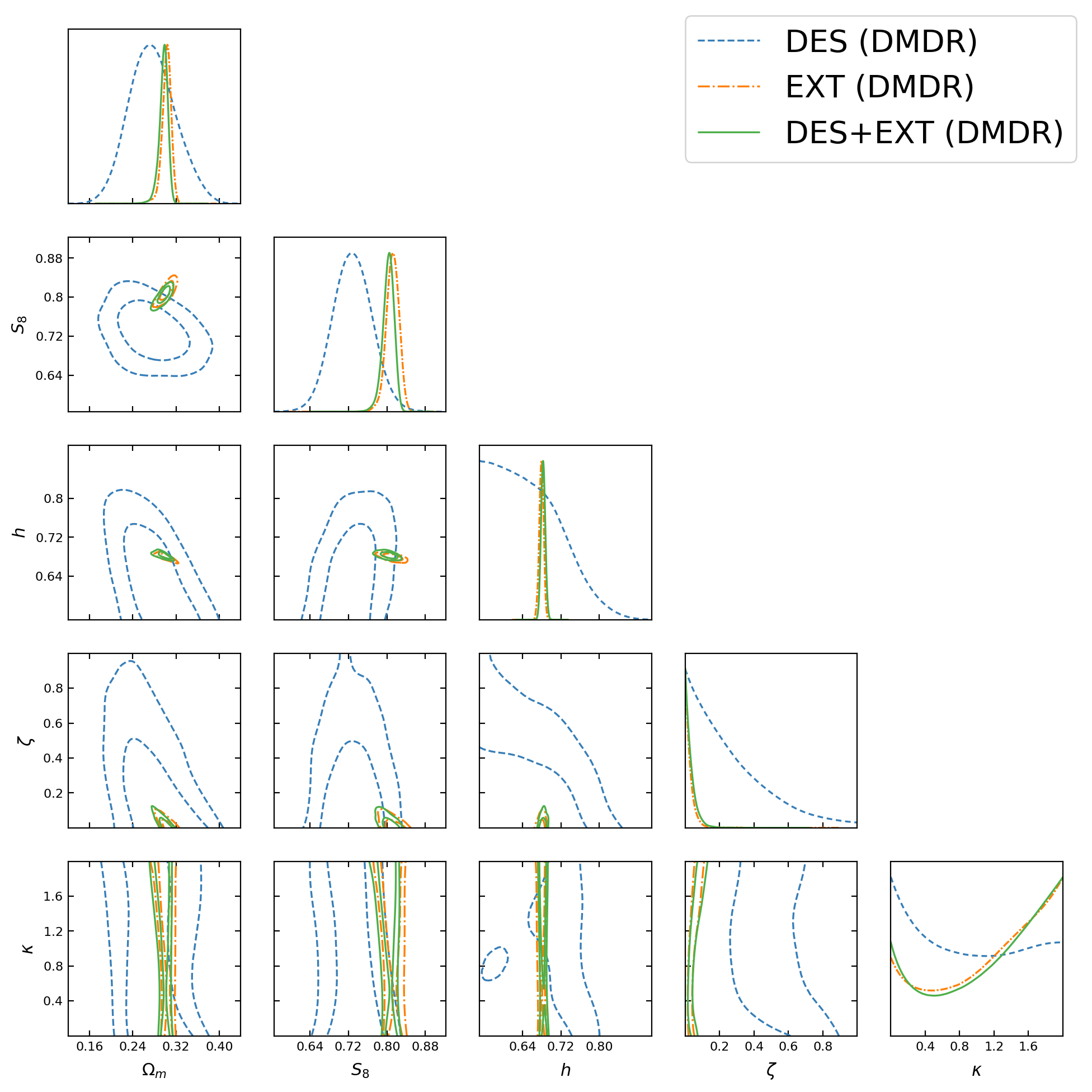}
    \caption{Constraints by DES-only, External-only, and DES+External data on the converted dark matter fraction $\zeta$ and rate $\kappa$, along with those on $\Omega_{\rm m}$, $S_8$, and $h$. }
    \label{fig:ddmpars}
\end{figure*}

\section{Results}
\label{results}

We now present our constraints on DMDR cosmology, followed by the tension and model-comparison results. 

\subsection{Constraints on DMDR model}

The constraints on DMDR parameters $\zeta$ and $\kappa$ are shown in figure~\ref{fig:ddmpars}, and their 1D marginalized statistics summarized in table~\ref{table:cosmoresults}. For the converted dark matter fraction $\zeta$, we find: 
\begin{alignat}{2}
    \zeta & < 0.32 & \quad \mbox{DES-only}\\[0.1cm]
          & < 0.030 & \quad \mbox{External-only} \\[0.1cm]
          & < 0.037 & \quad \mbox{DES+External.}
\end{alignat}

\renewcommand{\arraystretch}{1.5}
\begin{table*}[htbp]
\centering
\caption{1D marginalized statistics of cosmological parameters. The means of the marginalized 1D posteriors and $1 \sigma$ confidence levels are reported, with global maximum posterior sample in the parenthesis. The dashed lines mean that there is no constraint on the parameter (but we report the global posterior maximum), while the N/A means that the parameter is not relevant to the model studied. For the DES-only DMDR constraint, the global best fit of $\Omega_{\rm m}$ is about $2\sigma$ away from the mean value, possibly due to the $\zeta$--$\Omega_{\rm m}$ degeneracy. The degeneracy is broken for the External and DES+External datasets, when information from a wide redshift range is taken into consideration. }
\begin{tabular}{lccccc}
\hline
\rule[-0.7em]{0pt}{2em}
  & $h$ & $\Omega_{\rm m}$ & $S_8$ & $\zeta$ & $\kappa$\\
\hline
DES (DMDR) & $< 0.68 \; (0.64)  $ & $0.276^{+0.039}_{-0.046} \; ( 0.346)  $& $0.729\pm 0.040 \; (0.700) $ & $< 0.32 \; (0.01) $ & --- \; $(1.38)$\\
DES ($\Lambda$CDM) & $< 0.69 \;  (0.72) $ & $0.310^{+0.035}_{-0.040} \; (0.306) $& $0.726\pm 0.039 \; (0.723) $ & N/A & N/A \\
EXT (DMDR) & $0.6794\pm 0.0046 \; (0.6767) $ & $0.3025^{+0.0091}_{-0.0069} \; (0.3113) $& $0.812\pm 0.013 \; (0.829) $  & $< 0.030 \; (0.028)$ & --- \; $(0.0033) $\\
EXT ($\Lambda$CDM) & $0.6786\pm 0.0046 \; (0.6783) $ & $0.3085\pm 0.0059 \; (0.3093)$& $0.819\pm 0.011 \; (0.826) $ & N/A & N/A \\
DES+EXT (DMDR) & $0.6830\pm 0.0045 \; (0.6822)$ & $0.2970^{+0.0091}_{-0.0062} \; (0.2994) $& $0.803^{+0.013}_{-0.010} \; (0.808)  $ & $< 0.037 \; (0.020)$ & --- \; $(1.90) $ \\
DES+EXT ($\Lambda$CDM) & $0.6822\pm 0.0043 \; (0.6825)$ & $0.3038\pm 0.0054 \; (0.3036)$& $0.809^{+0.010}_{-0.009} \; (0.808)  $ & N/A & N/A \\
SH0ES &$0.740 \pm 0.014$ & N/A & N/A & N/A & N/A \\

\hline
\end{tabular}
\label{table:cosmoresults}
\end{table*}%

Note that we see a slightly looser constraint on $\zeta$ with DES+External dataset than External-only dataset. This is somewhat counter-intuitive, as our forecast predicted that weak lensing and galaxy clustering would tighten the constraint on $\zeta$ by anchoring the matter density at low redshift. However the Fisher forecast of course assumes Gaussian likelihood in all parameters. In the presence of non-Gaussianities, especially in a high-dimensional space, combined constraints are often (slightly) worse than those from individual probes. 

No constraint on conversion rate $\kappa$ is obtained;
see the bottom right of figure~\ref{fig:ddmpars}. This agrees with the expectation that $\kappa$ is unconstrained in the limit when the amount of converted dark matter, $\zeta$, is very small.

We can see a raising posterior profile towards the upper bound of the $\kappa$ prior. Although not statistically meaningful, such posterior profile suggest that we possibly underestimated the prior upper bound. Other explanations include the IA systematics and high-dimensional parameter space geometry. In any case, higher $\kappa$, namely even faster conversion that happens at extremely low-z is still open for investigation. However exploration of this avenue requires a more specific analysis, similar to one in models with a late dark-energy transition \cite{benevento2020can} in order to take the distance-ladder calibration into consideration. Hence we leave this for future work.

Other cosmological parameters that are of interest because they are tightly constrained or exhibit tensions between surveys --- $h$, $\Omega_{\rm m}$ and $S_8=\sigma_8 \sqrt{\Omega_{\rm m}/0.3}$ --- are also illustrated in the triangle plot figure~\ref{fig:ddmpars}, and summarized in table~\ref{table:cosmoresults}. 



\renewcommand{\arraystretch}{1.2}
\begin{table*}[tbp]
\centering
\caption{Difference in $\chi^2_{\rm MAP}$, evaluated at the maximum {\it a posteriori} point in parameter space, between DMDR and $\Lambda$CDM for different dataset combinations. }

\begin{tabular}{lcccccccc}
\hline
\rule[-0.7em]{0pt}{2em}
  & DES-Y1 3x2pt & Planck2018-CMB & Planck2018-lensing & Pantheon & 6dFGS & BOSS DR12 & MGS & Total\\
\hline
DES $\Delta \chi^2_{\rm MAP}$ & $-0.6$ & & & & & & & $-0.6$ \\
EXT $\Delta \chi^2_{\rm MAP}$ & & $0.0$ & $0.0$ & $0.1$ & $0.1$ & $0.7$ & $-0.1$ & $0.8$ \\
DES+EXT $\Delta \chi^2_{\rm MAP}$ & $0.7$ & $-0.4$ & $-0.4$ & $-0.0$ & $0.0$ & $0.3$ & $-0.1$ & $0.1$\\

\hline
\end{tabular}
\label{table:chisq}
\end{table*}%

\subsection{Model Comparison and Tensions}

As the tension between early and late universe surveys draws more and more attention in the cosmology community, there has been increasing number of works dedicated to quantify the concordance and discordance into statistical metrics \cite{raveri2019concordance, handley2019, wu2020hubble}. In this work, we quote Bayesian evidence and maximum a posteriori (MAP) $\chi^2$ difference as the model-comparison metrics, and use the "suspiciousness" metric defined in reference~\cite{handley2019}. We also report the one-dimensional differences in units of error bars for the parameters suspected to be tension, i.e. $h, \Omega_{\rm m}$ and $S_8$. We stress that 
we avoid combining any datasets that are known to be in tension, such as {\it Planck} and distance ladder (for $h$) or  {\it Planck} and DES  (for $S_8$).

We now report the model-comparison results.
\begin{itemize}
    \item \textbf{$\chi^2$ at MAP Cosmology}.
A very traditional criterion of the goodness of a model is the $\chi^2$ evaluated at the maximum a posteriori parameter values $\chi^2_{\rm MAP} = (d-M)^{\rm T} C^{-1} (d-M) |_{\rm MAP}$,
where $d$ is the full dataset, $M$ is the theory prediction evaluated at the maximum posterior sample, and $C$ is the covariance matrix of the full dataset. A preferred model should have smaller MAP $\chi^2$, and be punished by the number of extra parameters. Due to the non-gaussianity and the different normalization scheme of different survey likelihoods, we choose to report the effective $\chi^2$ defined as:
\begin{equation}
    \chi^2_{\rm MAP} = -2 \log{\mathcal{L}}|_{\rm MAP}.
\end{equation}

We ran an optimizer three times, adopting the {\fontfamily{qcr}\selectfont scipy} optimizer with Nelder–Mead method to calculate the MAP from the {\fontfamily{qcr}\selectfont
Polychord} chain samples; from these we report the best final MAP value. The $\chi^2$ difference between the DMDR and $\Lambda$CDM model is 
\begin{alignat}{2}
    \nonumber
    \Delta \chi^2_{\rm MAP} & = -0.6 & \quad \mbox{DES-only}\\
                                & = +0.8 & \quad \mbox{External-only} \\
                                & =+0.1 & \quad \mbox{DES+External}.
    \nonumber
\end{alignat}
as summarized in table~\ref{table:chisq}. Therefore our DMDR model does not give a substantially better global fit to the data than $\Lambda$CDM. 

\item \textbf{Bayesian Evidence Ratio.}
Bayesian evidence $\mathcal{Z}$ is defined as \begin{equation}
    \mathcal{Z} = \int \mathcal{L}(d|\theta) \Pi(\theta) d\theta
\end{equation}
where $\mathcal{L}$ is the likelihood, $d$ is the data vector, and $\theta$ are the model parameters. We report $\mathcal{Z}$ reported by the nested sampler {\fontfamily{qcr}\selectfont
Polychord}, with statistics done by {\fontfamily{qcr}\selectfont Anesthetic} \cite{anesthetic}.\footnote{https://github.com/williamjameshandley/anesthetic} The evidence ratio could be interpreted as the probability of two models given data through \cite{knuth2015bayesian}:
\begin{equation}
    \frac{P({\rm DMDR}|d,I)}{P({\rm \Lambda CDM}|d,I)} = \frac{P({\rm DMDR}|I)}{P({\rm \Lambda CDM}|I)} \frac{\mathcal{Z}({\rm DMDR})}{\mathcal{Z}({\rm \Lambda CDM})}
\end{equation}
where $I$ is the prior that these two models are in the consideration. Assuming no prior preference for either DMDR or $\Lambda$CDM, namely $P({\rm DMDR}|I) = P({\rm \Lambda CDM}|I)$, the ratio   of DMDR and $\Lambda$CDM probabilities  is equal to  the ratio of their respective evidences $\mathcal{Z}$. These, in turn, are reported by the {\fontfamily{qcr}\selectfont
Polychord} sampler; their ratio is
\begin{alignat}{2}
    \nonumber
     K = \frac{\mathcal{Z}({\rm DMDR})}{\mathcal{Z}({\rm \Lambda CDM})} & =  0.31 & \quad \mbox{DES-only}\\
                                & = 0.03 & \quad \mbox{External-only} \\
                                & =0.09 & \quad \mbox{DES+External.}
    \nonumber
\end{alignat}

We interpret the Bayesian evidence ratio in terms of the Jeffreys' scale (making this also consistent with DES-Y1 paper \cite{desy1}). Assuming an equal prior on $\Lambda$CDM and DMDR model, $0.31<K<1.0$ would indicate no conclusive preference for either model, $0.1<K<0.31$ would imply substantial evidence favouring $\Lambda$CDM, $0.031<K<0.1$ would imply strong evidence favouring $\Lambda$CDM, and $K<0.031$ would imply very strong evidence favouring $\Lambda$CDM \cite{jeffreys1961theory,robert2009harold}. 

Under Jeffreys' scale, our results therefore indicate that the DES-Y1-only dataset does not prefer either DMDR or $\Lambda$CDM, while the External-only dataset very strongly disfavors the DMDR model. Finally the combination of all datasets strongly disfavors DMDR.

\item \textbf{Suspiciousness.}
This tension statistic \cite{handley2019} has the merit of being less affected by the choice of the priors than Bayesian evidence. Suspiciousness $\mathcal{S}$ is defined in terms of the Bayesian evidence ratio $R$ and information ratio $I$:
\begin{equation}
\log \mathcal{S} = \log R- \log I, 
\end{equation}
where
\begin{alignat}{1}
    R & = \frac{\mathcal{Z}_{AB}}{\mathcal{Z}_A \mathcal{Z}_B}\\
    \log I & = \mathcal{D}_A + \mathcal{D}_B -\mathcal{D}_{AB}\\
    \mathcal{D} & = \int \mathcal{P}(\theta) \log \frac{\mathcal{P}(\theta)}{\Pi(\theta)} d \theta,
\end{alignat}
where $\mathcal{D}$ is the Kullback--Leibler divergence of the posterior against prior, quantifying the information gained by the data. The calculation of suspiciousness requires our knowledge of the posterior $\mathcal{P}$, prior $\Pi$, and evidence $\mathcal{Z}$ from MCMC chains. Here $A$ and $B$ stand for the DES-Y1 and External datasets  that we are comparing, and $AB$ for their combination. We report the $\log \mathcal{S}$ calculated by {\fontfamily{qcr}\selectfont Anesthetic} \cite{anesthetic}:
\begin{equation}
    \begin{aligned}
    \log \mathcal{S} & = -2.21,  \quad p = 0.08 & \quad \mbox{DMDR}\\
    \log \mathcal{S} & = -2.93, \quad p = 0.04 & \quad \mbox{$\Lambda$CDM}
    \end{aligned}
\end{equation}
where each $p$-value is interpreted as the probability that  datasets $A$ and $B$ can be both described by the parameters of the model. We therefore find that DMDR reduces the tension between DES and the external data, as indicated by a higher $p$-value, at the expense of two new parameters.

\begin{figure*}
\includegraphics[width=0.48\textwidth]{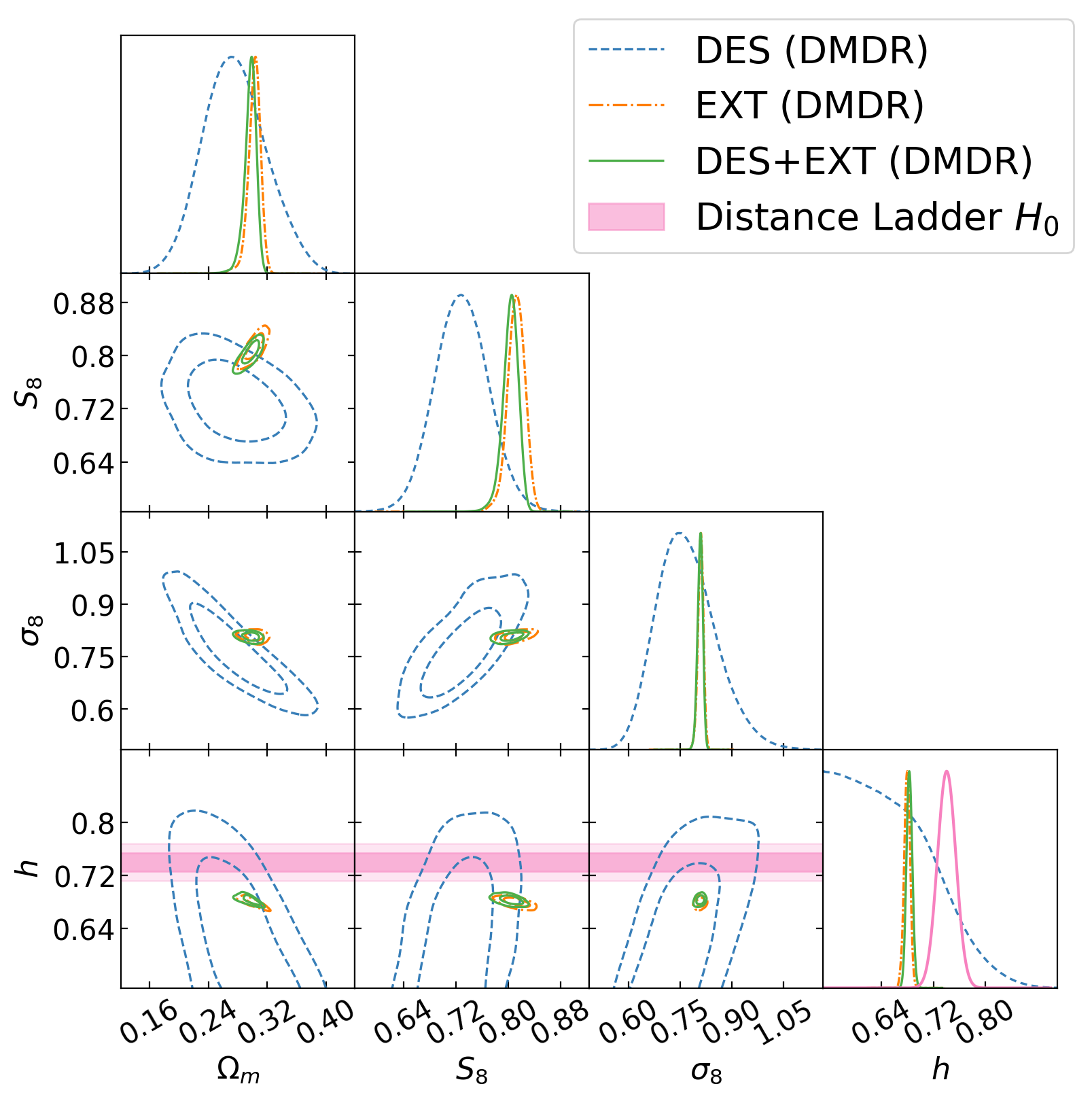}
\includegraphics[width=0.48\textwidth]{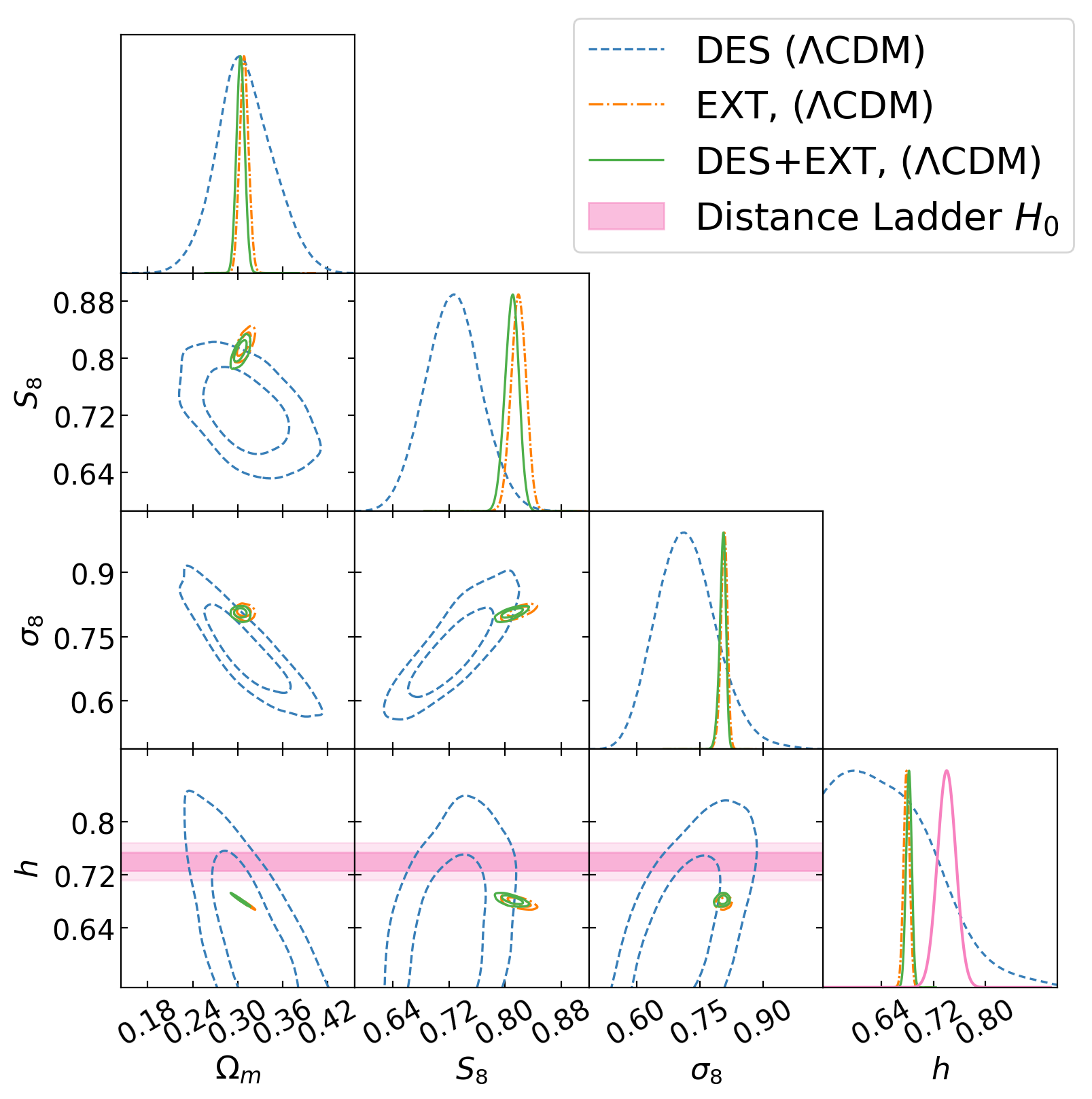}
\caption{Left panel: Cosmological parameters $\Omega_{\rm m}, S_8, \sigma_8, h$ constraints in DMDR model, reported for DES, External, and DES+External datasets, together with the local Hubble measurement \cite{riess2019large} in pink. Right panel: same plot in the $\Lambda$CDM cosmology. By comparing the panels involving $\sigma_8$, $S_8$ on both sides, we can see how DMDR reduced the tension in the matter density fields between DES and the CMB+Supernovae+BAOs. }
\label{fig:tensioncomparison}
\end{figure*}

\item \textbf{Hubble and $S_8$ tensions.}  
We now specifically investigate the impact of the new freedom in DMDR to two widely discussed tensions in $\Lambda$CDM: the $4$-$5 \sigma$ tension in the (scaled) Hubble constant $h$ between CMB and local measurements, and the $2$-$3 \sigma$ tension in $S_8$ between CMB and weak lensing plus clustering. We take the probability distribution of the parameter difference $\Delta h=h_{A}-h_{B}$ or, alternatively, $\Delta S_8=S_{8,A}-S_{8,B}$, from the 1D marginalized probability distribution obtained by different datasets. Here $A$ and $B$ are the two datasets between which we want to estimate the tension (in either $h$ or $S_8$). For a cosmological parameter of interest $\theta$, we integrate over the interval bounded by the $\Delta \theta$ values that have the equal posterior, and one of the boundaries is $\Delta \theta=0$. Thus we get the tension probability $p$:
\begin{equation}
    p = \int_{\Delta \theta=0}^{\rm eq-post} P(\Delta \theta = \theta_A-\theta_B) d\Delta \theta.
\end{equation}
We then interpret $p$ into $z-\sigma$ tension using
\begin{equation}
    p = {\rm erf}\left(\frac{z}{\sqrt{2}}\right).
\end{equation}
For the tension in the Hubble parameter, the dataset $A$ is the full DES+CMB+Supernovae+BAO data, while dataset $B$ is the Gaussian-distributed constraint on $h$ from the distance-ladder  measurement \cite{riess2019large}. For the $\Delta S_8$ tension, our $A$ dataset is the DES-Y1 3x2pt only data, while $B$ is the CMB+Supernovae+BAO External dataset. The zoomed-in constraints on $\Omega_m, S_8, \sigma_8$ and $h$ are illustrated in figure~\ref{fig:tensioncomparison}, over-plotted with the distance ladder measurement of $H_0$ from \cite{riess2019large}. We find that:

\begin{itemize}
\item When comparing the DES+External datasets with local Hubble measurement in \cite{riess2019large}, $h=0.7403\pm 0.0142$, the tension in $h$ assuming either DMDR or $\Lambda$CDM is $3.8 \sigma$. 

\item When comparing DES-Y1 dataset with External dataset, the tension in $S_8$ is $1.9 \sigma$ for DMDR model, slightly reduced from $2.3 \sigma$ for $\Lambda$CDM model.
\end{itemize}

Hence our DMDR model does not substantially alleviate the Hubble tension, but does help in reducing the $S_8$ tension.

\end{itemize}


\section{Conclusions}\label{sec:concl}

In this work, we test a late-time dark matter to dark radiation conversion model, dubbed the DMDR model, against cosmological data. Our model is specified by two new parameters defined in equations~(\ref{eq:background_1}) and (\ref{eq:background_2}): the fraction of dark matter that has converted $\zeta$, and the rate of its conversion (to dark radiation) $\kappa$.
We work out the perturbation equations in this model, and incorporate them in the Einstein--Boltzmann code CAMB \cite{camb}.
Our analysis pipeline is modified for the DMDR model in the following respects. 
1) we scale-dependently correct the shear and intrinsic alignment terms in the two-point correlation functions to account for the non-trivial relation between gravitational field and matter density perturbation field, and 2) we adopt conservative scale cuts to protect the analysis against systematic errors due to the modeling of clustering on nonlinear scales. In our analysis, we principally consider  the  DES-Y1 "3x2pt" (weak lensing and galaxy clustering) data. We also study the impact of adding external datasets: Planck-2018 CMB power spectra (TT, TE, EE, and lensing spectrum); Pantheon compilation of type Ia supernovae data; and compressed BAO measurements from BOSS-DR12, MGS and 6dFGS surveys.  

The constraint on the fraction of the converted dark matter obtained from all data combined is $\zeta < 0.037$. We find no constraint on the conversion rate parameter $\kappa$ as expected in the limit when $\zeta\rightarrow 0$. We further find that the evidence-ratio test applied with the full combined data does not favor the DMDR model compared to $\Lambda$CDM. DMDR does however reduce the suspiciousness tension metric between DES-Y1 and the combination of CMB, Supernovae and BAO data, raising the probability that DES and external data are concordant from 4\% to 8\%. 
Finally, DMDR  does not help in alleviating the Hubble tension, but does reduce the tension in the DES and external-data measurements of $S_8=\sigma_8 \sqrt{\Omega_{\rm m}/0.3}$, making it go from $2.3 \sigma$ (in $\Lambda$CDM) to $1.9 \sigma$ (in DMDR). 

We stress that the above conclusions are drawn for the late-universe dark matter-dark radiation conversion model introduced in section \ref{sec:backgroundeqs}. Further generalizations of this catalogue \cite{doroshkevich1989large,oguri2003decaying,bjaelde2012origin,wang2012effects,cirelli2012gamma,wang2013lyman,blackadder2014dark,aoyama2014evolution,enqvist2015,poulin2016,bringmann,pandey2019alleviating,vattis2019late,archidiacono2019constraining, haridasu2020late,enqvist2020constraints,clark2020cmb}, for example where dark matter is a composition of some fraction of interacting dark matter and cold dark matter, or where the transition time is short, or the transition occurs in the early universe, were not considered in this work. These variants could in principle better fit the background evolution of the universe than the model we studied, and are thus a promising target for further investigations. 


There are several other directions in which our analysis could be extended. One possibility is to model the nonlinear matter power spectrum in real and redshift space in DMDR models \cite{cataneo2019road,giblin2019road,d2020cosmological}. This could be particularly helpful for DES year-3 and year-6 data which have more statistical power and where pushing to smaller, nonlinear scales could improve the constraints. Another future direction is to enable the use of the uncompressed BAO data (that is, the broadband galaxy and quasar power spectra). This would potentially improve the constraints for  not only the DMDR model but also other beyond-$\Lambda$CDM models, and could become an important analysis tool for future surveys such as those to be undertaken by DESI, the Rubin Observatory (LSST), Euclid, and the Roman Space Telescope. 

Our investigation was limited to galaxy clustering, weak lensing, and galaxy-galaxy lensing which are united in the so-called "3x2" analysis. Recent years have seen the emergence of new, promising cosmological probes which, when incorporated, could improve the constraints presented here. For example, the Lyman-$\alpha$ BAO measurements from high-redshift quasars and clustering obtained from the 21-cm signal could both be very helpful for constraining DMDR-type models where slow transition happen between $z\sim 1$ and recombination. The medium redshift measurements can fill in the blank in the current cosmological observations concentrated on two ends of the time stretch. It will be exciting to see if incorporating new cosmological probes and combining them with the improved 3x2 analyses from Stage IV dark-energy surveys can help shed light on DMDR-type models.

\section*{Acknowledgement}
We thank Prof. Shanjie Zhang, Prof. Jianming Jin and Dr. Robert C. Forrey for the development of the Hypergeometric Function calculation routine used in our DMDR-CAMB.

Funding for the DES Projects has been provided by the U.S. Department of Energy, the U.S. National Science Foundation, the Ministry of Science and Education of Spain, 
the Science and Technology Facilities Council of the United Kingdom, the Higher Education Funding Council for England, the National Center for Supercomputing Applications at the University of Illinois at Urbana-Champaign, the Kavli Institute of Cosmological Physics at the University of Chicago, 
the Center for Cosmology and Astro-Particle Physics at the Ohio State University,
the Mitchell Institute for Fundamental Physics and Astronomy at Texas A\&M University, Financiadora de Estudos e Projetos, 
Funda{\c c}{\~a}o Carlos Chagas Filho de Amparo {\`a} Pesquisa do Estado do Rio de Janeiro, Conselho Nacional de Desenvolvimento Cient{\'i}fico e Tecnol{\'o}gico and 
the Minist{\'e}rio da Ci{\^e}ncia, Tecnologia e Inova{\c c}{\~a}o, the Deutsche Forschungsgemeinschaft and the Collaborating Institutions in the Dark Energy Survey. 

The Collaborating Institutions are Argonne National Laboratory, the University of California at Santa Cruz, the University of Cambridge, Centro de Investigaciones Energ{\'e}ticas, 
Medioambientales y Tecnol{\'o}gicas-Madrid, the University of Chicago, University College London, the DES-Brazil Consortium, the University of Edinburgh, 
the Eidgen{\"o}ssische Technische Hochschule (ETH) Z{\"u}rich, 
Fermi National Accelerator Laboratory, the University of Illinois at Urbana-Champaign, the Institut de Ci{\`e}ncies de l'Espai (IEEC/CSIC), 
the Institut de F{\'i}sica d'Altes Energies, Lawrence Berkeley National Laboratory, the Ludwig-Maximilians Universit{\"a}t M{\"u}nchen and the associated Excellence Cluster Universe, 
the University of Michigan, NFS's NOIRLab, the University of Nottingham, The Ohio State University, the University of Pennsylvania, the University of Portsmouth, 
SLAC National Accelerator Laboratory, Stanford University, the University of Sussex, Texas A\&M University, and the OzDES Membership Consortium.

Based in part on observations at Cerro Tololo Inter-American Observatory at NSF's NOIRLab (NOIRLab Prop. ID 2012B-0001; PI: J. Frieman), which is managed by the Association of Universities for Research in Astronomy (AURA) under a cooperative agreement with the National Science Foundation.

The DES data management system is supported by the National Science Foundation under Grant Numbers AST-1138766 and AST-1536171.
The DES participants from Spanish institutions are partially supported by MICINN under grants ESP2017-89838, PGC2018-094773, PGC2018-102021, SEV-2016-0588, SEV-2016-0597, and MDM-2015-0509, some of which include ERDF funds from the European Union. IFAE is partially funded by the CERCA program of the Generalitat de Catalunya.
Research leading to these results has received funding from the European Research
Council under the European Union's Seventh Framework Program (FP7/2007-2013) including ERC grant agreements 240672, 291329, and 306478.
We  acknowledge support from the Brazilian Instituto Nacional de Ci\^encia
e Tecnologia (INCT) do e-Universo (CNPq grant 465376/2014-2).

This manuscript has been authored by Fermi Research Alliance, LLC under Contract No. DE-AC02-07CH11359 with the U.S. Department of Energy, Office of Science, Office of High Energy Physics.

\bibliographystyle{ieeetr}
\bibliography{ddmdraft}

\begin{thebibliography}{100}

\bibitem{planck2018}
N.~Aghanim, Y.~Akrami, M.~Ashdown, J.~Aumont, C.~Baccigalupi, M.~Ballardini,
  A.~Banday, R.~Barreiro, N.~Bartolo, S.~Basak, {\em et~al.}, ``Planck 2018
  results. vi. cosmological parameters,'' {\em arXiv preprint
  arXiv:1807.06209}, 2018.

\bibitem{verdetensions}
L.~Verde, T.~Treu, and A.~G. Riess, ``Tensions between the early and late
  universe,'' {\em Nature Astronomy}, vol.~3, pp.~891--895, 2019.

\bibitem{Riess:2020sih}
A.~G. Riess, ``The expansion of the universe is faster than expected,'' {\em
  Nature Reviews Physics}, vol.~2, no.~1, pp.~10--12, 2020.

\bibitem{huang2018near}
C.~D. Huang, A.~G. Riess, S.~L. Hoffmann, C.~Klein, J.~Bloom, W.~Yuan, L.~M.
  Macri, D.~O. Jones, P.~A. Whitelock, S.~Casertano, {\em et~al.}, ``A
  near-infrared period--luminosity relation for miras in ngc 4258, an anchor
  for a new distance ladder,'' {\em The Astrophysical Journal}, vol.~857,
  no.~1, p.~67, 2018.

\bibitem{ariess}
A.~G. Riess, S.~Casertano, W.~Yuan, L.~M. Macri, and D.~Scolnic, ``Large
  magellanic cloud cepheid standards provide a 1\% foundation for the
  determination of the hubble constant and stronger evidence for physics beyond
  $\lambda$cdm,'' {\em The Astrophysical Journal}, vol.~876, no.~1, p.~85,
  2019.

\bibitem{wendy}
W.~L. Freedman, B.~F. Madore, D.~Hatt, T.~J. Hoyt, I.~S. Jang, R.~L. Beaton,
  C.~R. Burns, M.~G. Lee, A.~J. Monson, J.~R. Neeley, {\em et~al.}, ``The
  carnegie-chicago hubble program. viii. an independent determination of the
  hubble constant based on the tip of the red giant branch,'' {\em The
  Astrophysical Journal}, vol.~882, no.~1, p.~34, 2019.

\bibitem{h0licow}
K.~C. Wong, S.~H. Suyu, G.~C.-F. Chen, C.~E. Rusu, M.~Millon, D.~Sluse,
  V.~Bonvin, C.~D. Fassnacht, S.~Taubenberger, M.~W. Auger, {\em et~al.},
  ``H0licow xiii. a 2.4\% measurement of $ h_ {0} $ from lensed quasars:
  $5.3\backslash \sigma$ tension between early and late-universe probes,'' {\em
  arXiv preprint arXiv:1907.04869}, 2019.

\bibitem{DiValentino:2019ffd}
E.~Di~Valentinoa, A.~Melchiorrib, O.~Menac, and S.~Vagnozzid, ``Interacting
  dark energy in the early 2020s: a promising solution to the h0 and cosmic
  shear tensions,'' {\em arXiv preprint arXiv:1908.04281}, 2019.

\bibitem{poulin2019early}
V.~Poulin, T.~L. Smith, T.~Karwal, and M.~Kamionkowski, ``Early dark energy can
  resolve the hubble tension,'' {\em Physical review letters}, vol.~122,
  no.~22, p.~221301, 2019.

\bibitem{lin2019acoustic}
M.-X. Lin, G.~Benevento, W.~Hu, and M.~Raveri, ``Acoustic dark energy:
  Potential conversion of the hubble tension,'' {\em Physical Review D},
  vol.~100, no.~6, p.~063542, 2019.

\bibitem{blinov2020warm}
N.~Blinov, C.~Keith, and D.~Hooper, ``Warm decaying dark matter and the hubble
  tension,'' {\em Journal of Cosmology and Astroparticle Physics}, vol.~2020,
  no.~06, p.~005, 2020.

\bibitem{sola2020brans}
J.~Sola, A.~Gomez-Valent, J.~d.~C. Perez, and C.~Moreno-Pulido, ``Brans-dicke
  cosmology with a $\lambda$-term: a possible solution to $\lambda$cdm
  tensions,'' {\em arXiv preprint arXiv:2006.04273}, 2020.

\bibitem{jedamzik2020relieving}
K.~Jedamzik and L.~Pogosian, ``Relieving the hubble tension with primordial
  magnetic fields,'' {\em arXiv preprint arXiv:2004.09487}, 2020.

\bibitem{elizalde2020analysis}
E.~Elizalde, M.~Khurshudyan, S.~D. Odintsov, and R.~Myrzakulov, ``An analysis
  of the $ h\_ $\{$0$\}$ $ tension problem in a universe with a viscous dark
  fluid,'' {\em arXiv preprint arXiv:2006.01879}, 2020.

\bibitem{li2020generalised}
X.~Li and A.~Shafieloo, ``Generalised emergent dark energy model: Confronting
  $lambda$ and pede,'' {\em arXiv preprint arXiv:2001.05103}, 2020.

\bibitem{Yang:2020myd}
W.~Yang, E.~Di~Valentino, S.~Pan, and O.~Mena, ``{A complete model of
  Phenomenologically Emergent Dark Energy},'' {\em arXiv preprint
  arXiv:2007.02927}, 2020.

\bibitem{hart2020updated}
L.~Hart and J.~Chluba, ``Updated fundamental constant constraints from planck
  2018 data and possible relations to the hubble tension,'' {\em Monthly
  Notices of the Royal Astronomical Society}, vol.~493, no.~3, pp.~3255--3263,
  2020.

\bibitem{ballardini2020scalar}
M.~Ballardini, M.~Braglia, F.~Finelli, D.~Paoletti, A.~A. Starobinsky, and
  C.~Umilt{\`a}, ``Scalar-tensor theories of gravity, neutrino physics, and the
  $ h\_0 $ tension,'' {\em arXiv preprint arXiv:2004.14349}, 2020.

\bibitem{kids2016}
H.~Hildebrandt, M.~Viola, C.~Heymans, S.~Joudaki, K.~Kuijken, C.~Blake,
  T.~Erben, B.~Joachimi, D.~Klaes, L.~t. Miller, {\em et~al.}, ``Kids-450:
  Cosmological parameter constraints from tomographic weak gravitational
  lensing,'' {\em Monthly Notices of the Royal Astronomical Society}, vol.~465,
  no.~2, pp.~1454--1498, 2016.

\bibitem{leauthaud2017lensing}
A.~Leauthaud, S.~Saito, S.~Hilbert, A.~Barreira, S.~More, M.~White, S.~Alam,
  P.~Behroozi, K.~Bundy, J.~Coupon, {\em et~al.}, ``Lensing is low: cosmology,
  galaxy formation or new physics?,'' {\em Monthly Notices of the Royal
  Astronomical Society}, vol.~467, no.~3, pp.~3024--3047, 2017.

\bibitem{lin2017cosmological}
W.~Lin and M.~Ishak, ``Cosmological discordances. ii. hubble constant, planck
  and large-scale-structure data sets,'' {\em Physical Review D}, vol.~96,
  no.~8, p.~083532, 2017.

\bibitem{desy1}
T.~Abbott, F.~Abdalla, A.~Alarcon, J.~Aleksi{\'c}, S.~Allam, S.~Allen,
  A.~Amara, J.~Annis, J.~Asorey, S.~Avila, {\em et~al.}, ``Dark energy survey
  year 1 results: Cosmological constraints from galaxy clustering and weak
  lensing,'' {\em Physical Review D}, vol.~98, no.~4, p.~043526, 2018.

\bibitem{DiValentino:2018gcu}
E.~Di~Valentino and S.~Bridle, ``Exploring the tension between current cosmic
  microwave background and cosmic shear data,'' {\em Symmetry}, vol.~10,
  no.~11, p.~585, 2018.

\bibitem{2019kidscosebi}
M.~Asgari, T.~Tr{\"o}ster, C.~Heymans, H.~Hildebrandt, J.~L. v.~d. Busch, A.~H.
  Wright, A.~Choi, T.~Erben, B.~Joachimi, S.~Joudaki, {\em et~al.}, ``Kids+
  viking-450 and des-y1 combined: Mitigating baryon feedback uncertainty with
  cosebis,'' {\em arXiv preprint arXiv:1910.05336}, 2019.

\bibitem{buen2015non}
M.~A. Buen-Abad, G.~Marques-Tavares, and M.~Schmaltz, ``Non-abelian dark matter
  and dark radiation,'' {\em Physical Review D}, vol.~92, no.~2, p.~023531,
  2015.

\bibitem{murgia2016constraints}
R.~Murgia, S.~Gariazzo, and N.~Fornengo, ``Constraints on the coupling between
  dark energy and dark matter from cmb data,'' {\em Journal of Cosmology and
  Astroparticle Physics}, vol.~2016, no.~04, p.~014, 2016.

\bibitem{di2018reducing}
E.~Di~Valentino, C.~B{\oe}hm, E.~Hivon, and F.~R. Bouchet, ``Reducing the h 0
  and $\sigma$ 8 tensions with dark matter-neutrino interactions,'' {\em
  Physical Review D}, vol.~97, no.~4, p.~043513, 2018.

\bibitem{hill2020early}
J.~C. Hill, E.~McDonough, M.~W. Toomey, and S.~Alexander, ``Early dark energy
  does not restore cosmological concordance,'' {\em arXiv preprint
  arXiv:2003.07355}, 2020.

\bibitem{ivanov2020constraining}
M.~M. Ivanov, E.~McDonough, J.~C. Hill, M.~Simonovi{\'c}, M.~W. Toomey,
  S.~Alexander, and M.~Zaldarriaga, ``Constraining early dark energy with
  large-scale structure,'' {\em arXiv preprint arXiv:2006.11235}, 2020.

\bibitem{klypin2020clustering}
A.~Klypin, V.~Poulin, F.~Prada, J.~Primack, M.~Kamionkowski, V.~Avila-Reese,
  A.~Rodriguez-Puebla, P.~Behroozi, D.~Hellinger, and T.~L. Smith, ``Clustering
  and halo abundances in early dark energy cosmological models,'' {\em arXiv
  preprint arXiv:2006.14910}, 2020.

\bibitem{doroshkevich1989large}
A.~Doroshkevich, A.~Klypin, and M.~Khlopov, ``Large-scale structure of the
  universe in unstable dark matter models,'' {\em Monthly Notices of the Royal
  Astronomical Society}, vol.~239, no.~3, pp.~923--938, 1989.

\bibitem{oguri2003decaying}
M.~Oguri, K.~Takahashi, H.~Ohno, and K.~Kotake, ``Decaying cold dark matter and
  the evolution of the cluster abundance,'' {\em The Astrophysical Journal},
  vol.~597, no.~2, p.~645, 2003.

\bibitem{Wang:2010ma}
M.-Y. Wang and A.~R. Zentner, ``{Weak Gravitational Lensing as a Method to
  Constrain Unstable Dark Matter},'' {\em Phys. Rev. D}, vol.~82, p.~123507,
  2010.

\bibitem{cirelli2012gamma}
M.~Cirelli, E.~Moulin, P.~Panci, P.~D. Serpico, and A.~Viana, ``Gamma ray
  constraints on decaying dark matter,'' {\em Physical Review D}, vol.~86,
  no.~8, p.~083506, 2012.

\bibitem{wang2012effects}
M.-Y. Wang and A.~R. Zentner, ``Effects of unstable dark matter on large-scale
  structure and constraints from future surveys,'' {\em Physical Review D},
  vol.~85, no.~4, p.~043514, 2012.

\bibitem{bjaelde2012origin}
O.~E. Bjaelde, S.~Das, and A.~Moss, ``Origin of $\delta$neff as a result of an
  interaction between dark radiation and dark matter,'' {\em Journal of
  Cosmology and Astroparticle Physics}, vol.~2012, no.~10, p.~017, 2012.

\bibitem{wang2013lyman}
M.-Y. Wang, R.~A. Croft, A.~H. Peter, A.~R. Zentner, and C.~W. Purcell,
  ``Lyman-$\alpha$ forest constraints on decaying dark matter,'' {\em Physical
  Review D}, vol.~88, no.~12, p.~123515, 2013.

\bibitem{blackadder2014dark}
G.~Blackadder and S.~M. Koushiappas, ``Dark matter with two-and many-body
  decays and supernovae type ia,'' {\em Physical Review D}, vol.~90, no.~10,
  p.~103527, 2014.

\bibitem{aoyama2014evolution}
S.~Aoyama, T.~Sekiguchi, K.~Ichiki, and N.~Sugiyama, ``Evolution of
  perturbations and cosmological constraints in decaying dark matter models
  with arbitrary decay mass products,'' {\em Journal of Cosmology and
  Astroparticle Physics}, vol.~2014, no.~07, p.~021, 2014.

\bibitem{enqvist2015}
K.~Enqvist, S.~Nadathur, T.~Sekiguchi, and T.~Takahashi, ``Decaying dark matter
  and the tension in $\sigma$8,'' {\em Journal of Cosmology and Astroparticle
  Physics}, vol.~2015, no.~09, p.~067, 2015.

\bibitem{poulin2016}
V.~Poulin, S.~Pasquale, D., and L.~Julien, ``{A fresh look at linear
  cosmological constraints on a decaying dark matter component},'' {\em Journal
  of Cosmology and Astroparticle Physics}, vol.~08, 2016.

\bibitem{bringmann}
T.~Bringmann, F.~Kahlhoefer, K.~Schmidt-Hoberg, and P.~Walia, ``Converting
  nonrelativistic dark matter to radiation,'' {\em Physical Review D}, vol.~98,
  no.~2, p.~023543, 2018.

\bibitem{pandey2019alleviating}
K.~L. Pandey, T.~Karwal, and S.~Das, ``Alleviating the h0 and s8 anomalies with
  a decaying dark matter model,'' {\em Journal of Cosmology and Astroparticle
  Physics}, 2019.

\bibitem{vattis2019late}
K.~Vattis, S.~M. Koushiappas, and A.~Loeb, ``Late universe decaying dark matter
  can relieve the h\_0 tension,'' {\em arXiv preprint arXiv:1903.06220}, 2019.

\bibitem{archidiacono2019constraining}
M.~Archidiacono, D.~C. Hooper, R.~Murgia, S.~Bohr, J.~Lesgourgues, and M.~Viel,
  ``Constraining dark matter-dark radiation interactions with cmb, bao, and
  lyman-$\alpha$,'' {\em Journal of Cosmology and Astroparticle Physics},
  vol.~2019, no.~10, p.~055, 2019.

\bibitem{clark2020cmb}
S.~J. Clark, K.~Vattis, and S.~M. Koushiappas, ``Cmb constraints on
  late-universe decaying dark matter as a solution to the $ h\_0 $ tension,''
  {\em arXiv preprint arXiv:2006.03678}, 2020.

\bibitem{haridasu2020late}
B.~S. Haridasu and M.~Viel, ``Late-time decaying dark matter: constraints and
  implications for the $ h\_0 $-tension,'' {\em arXiv preprint
  arXiv:2004.07709}, 2020.

\bibitem{enqvist2020constraints}
K.~Enqvist, S.~Nadathur, T.~Sekiguchi, and T.~Takahashi, ``Constraints on
  decaying dark matter from weak lensing and cluster counts,'' {\em Journal of
  Cosmology and Astroparticle Physics}, vol.~2020, no.~04, p.~015, 2020.

\bibitem{tulin2018dark}
S.~Tulin and H.-B. Yu, ``Dark matter self-interactions and small scale
  structure,'' {\em Physics Reports}, vol.~730, pp.~1--57, 2018.

\bibitem{valli2018dark}
M.~Valli and H.-B. Yu, ``Dark matter self-interactions from the internal
  dynamics of dwarf spheroidals,'' {\em Nature Astronomy}, vol.~2, no.~11,
  pp.~907--912, 2018.

\bibitem{smallscalechallenges}
J.~S. Bullock and M.~Boylan-Kolchin, ``Small-scale challenges to the
  $\lambda$cdm paradigm,'' {\em Annual Review of Astronomy and Astrophysics},
  vol.~55, 2017.

\bibitem{aiola2015gaussian}
S.~Aiola, A.~Kosowsky, and B.~Wang, ``Gaussian approximation of peak values in
  the integrated sachs-wolfe effect,'' {\em Physical Review D}, vol.~91, no.~4,
  p.~043510, 2015.

\bibitem{kovacs2019more}
A.~Kov{\'a}cs, C.~S{\'a}nchez, J.~Garc{\'\i}a-Bellido, J.~Elvin-Poole,
  N.~Hamaus, V.~Miranda, S.~Nadathur, T.~Abbott, F.~Abdalla, J.~Annis, {\em
  et~al.}, ``More out of less: an excess integrated sachs--wolfe signal from
  supervoids mapped out by the dark energy survey,'' {\em Monthly Notices of
  the Royal Astronomical Society}, vol.~484, no.~4, pp.~5267--5277, 2019.

\bibitem{aguilar2019towards}
M.~Aguilar, L.~A. Cavasonza, B.~Alpat, G.~Ambrosi, L.~Arruda, N.~Attig,
  P.~Azzarello, A.~Bachlechner, F.~Barao, A.~Barrau, {\em et~al.}, ``Towards
  understanding the origin of cosmic-ray electrons,'' {\em Physical review
  letters}, vol.~122, no.~10, p.~101101, 2019.

\bibitem{boyarsky2014unidentified}
A.~Boyarsky, O.~Ruchayskiy, D.~Iakubovskyi, and J.~Franse, ``Unidentified line
  in x-ray spectra of the andromeda galaxy and perseus galaxy cluster,'' {\em
  Physical review letters}, vol.~113, no.~25, p.~251301, 2014.

\bibitem{wang2014cosmological}
M.-Y. Wang, A.~H. Peter, L.~E. Strigari, A.~R. Zentner, B.~Arant,
  S.~Garrison-Kimmel, and M.~Rocha, ``Cosmological simulations of decaying dark
  matter: implications for small-scale structure of dark matter haloes,'' {\em
  Monthly Notices of the Royal Astronomical Society}, vol.~445, no.~1,
  pp.~614--629, 2014.

\bibitem{abazajian2017sterile}
K.~N. Abazajian, ``Sterile neutrinos in cosmology,'' {\em Physics Reports},
  vol.~711, pp.~1--28, 2017.

\bibitem{farzan2019dark}
Y.~Farzan and M.~Rajaee, ``Dark matter decaying into millicharged particles as
  a solution to ams-02 positron excess,'' {\em Journal of Cosmology and
  Astroparticle Physics}, vol.~2019, no.~04, p.~040, 2019.

\bibitem{das2020galactic}
A.~Das, B.~Dasgupta, and A.~Ray, ``Galactic positron excess from selectively
  enhanced dark matter annihilation,'' {\em Physical Review D}, vol.~101,
  no.~6, p.~063014, 2020.

\bibitem{ishiwata2020probing}
K.~Ishiwata, O.~Macias, S.~Ando, and M.~Arimoto, ``Probing heavy dark matter
  decays with multi-messenger astrophysical data,'' {\em Journal of Cosmology
  and Astroparticle Physics}, vol.~2020, no.~01, p.~003, 2020.

\bibitem{dessert2020dark}
C.~Dessert, N.~L. Rodd, and B.~R. Safdi, ``The dark matter interpretation of
  the 3.5-kev line is inconsistent with blank-sky observations,'' {\em
  Science}, vol.~367, no.~6485, pp.~1465--1467, 2020.

\bibitem{bhargava2020xmm}
S.~Bhargava, P.~Giles, A.~Romer, T.~Jeltema, J.~Mayers, A.~Bermeo, M.~Hilton,
  R.~Wilkinson, C.~Vergara, C.~Collins, {\em et~al.}, ``The xmm cluster survey:
  new evidence for the 3.5 kev feature in clusters is inconsistent with a dark
  matter origin,'' {\em Monthly Notices of the Royal Astronomical Society},
  2020.

\bibitem{Wang:2015fia}
M.-Y. Wang, L.~E. Strigari, M.~R. Lovell, C.~S. Frenk, and A.~R. Zentner,
  ``{Mass assembly history and infall time of the Fornax dwarf spheroidal
  galaxy},'' {\em Mon. Not. Roy. Astron. Soc.}, vol.~457, no.~4,
  pp.~4248--4261, 2016.

\bibitem{pospelov2009r}
M.~Pospelov and M.~Trott, ``R-parity preserving super-wimp decays,'' {\em
  Journal of High Energy Physics}, vol.~2009, no.~04, p.~044, 2009.

\bibitem{allahverdi2015dark}
R.~Allahverdi, B.~Dutta, F.~S. Queiroz, L.~E. Strigari, and M.-Y. Wang, ``Dark
  matter from late invisible decays to and of gravitinos,'' {\em Physical
  Review D}, vol.~91, no.~5, p.~055033, 2015.

\bibitem{ulrich}
U.~Ellwanger, H.~Cyril, and T.~Ana, M., ``{ The next-to-minimal supersymmetric
  standard model.},'' {\em Physics Reports}, vol.~496, no.~1, 2010.

\bibitem{higaki2012dark}
T.~Higaki and F.~Takahashi, ``Dark radiation and dark matter in large volume
  compactifications,'' {\em Journal of High Energy Physics}, vol.~2012, no.~11,
  p.~125, 2012.

\bibitem{edges}
J.~D. Bowman, A.~E. Rogers, R.~A. Monsalve, T.~J. Mozdzen, and N.~Mahesh, ``An
  absorption profile centred at 78 megahertz in the sky-averaged spectrum,''
  {\em Nature}, vol.~555, no.~7694, p.~67, 2018.

\bibitem{bondarenko2020constraining}
K.~Bondarenko, J.~Pradler, and A.~Sokolenko, ``Constraining dark photons and
  their connection to 21 cm cosmology with cmb data,'' {\em arXiv preprint
  arXiv:2002.08942}, 2020.

\bibitem{carr2016}
C.~Bernard, K.~Florian, and S.~Marit, ``{ Primordial black holes as dark
  matter.},'' {\em PHYSICAL REVIEW D}, vol.~94, no.~8, 2016.

\bibitem{raidal2017gravitational}
M.~Raidal, V.~Vaskonen, and H.~Veerm{\"a}e, ``Gravitational waves from
  primordial black hole mergers,'' {\em Journal of Cosmology and Astroparticle
  Physics}, vol.~2017, no.~09, p.~037, 2017.

\bibitem{masina2020dark}
I.~Masina, ``Dark matter and dark radiation from evaporating primordial black
  holes,'' {\em arXiv preprint arXiv:2004.04740}, 2020.

\bibitem{laha2019primordial}
R.~Laha, ``Primordial black holes as a dark matter candidate are severely
  constrained by the galactic center 511 kev $\gamma$-ray line,'' {\em Physical
  Review Letters}, vol.~123, no.~25, p.~251101, 2019.

\bibitem{clesse2018seven}
S.~Clesse and J.~Garc{\'\i}a-Bellido, ``Seven hints for primordial black hole
  dark matter,'' {\em Physics of the Dark Universe}, vol.~22, pp.~137--146,
  2018.

\bibitem{montero2019revisiting}
P.~Montero-Camacho, X.~Fang, G.~Vasquez, M.~Silva, and C.~M. Hirata,
  ``Revisiting constraints on asteroid-mass primordial black holes as dark
  matter candidates,'' {\em Journal of Cosmology and Astroparticle Physics},
  vol.~2019, no.~08, p.~031, 2019.

\bibitem{smyth2020updated}
N.~Smyth, S.~Profumo, S.~English, T.~Jeltema, K.~McKinnon, and P.~Guhathakurta,
  ``Updated constraints on asteroid-mass primordial black holes as dark
  matter,'' {\em Physical Review D}, vol.~101, no.~6, p.~063005, 2020.

\bibitem{jones2018measuring}
D.~Jones, D.~Scolnic, A.~Riess, A.~Rest, R.~Kirshner, E.~Berger, R.~Kessler,
  Y.-C. Pan, R.~Foley, R.~Chornock, {\em et~al.}, ``Measuring dark energy
  properties with photometrically classified pan-starrs supernovae. ii.
  cosmological parameters,'' {\em The Astrophysical Journal}, vol.~857, no.~1,
  p.~51, 2018.

\bibitem{bossdata}
S.~Alam, M.~Ata, S.~Bailey, F.~Beutler, D.~Bizyaev, J.~A. Blazek, A.~S. Bolton,
  J.~R. Brownstein, A.~Burden, C.-H. Chuang, {\em et~al.}, ``The clustering of
  galaxies in the completed sdss-iii baryon oscillation spectroscopic survey:
  cosmological analysis of the dr12 galaxy sample,'' {\em Monthly Notices of
  the Royal Astronomical Society}, vol.~470, no.~3, pp.~2617--2652, 2017.

\bibitem{mgs}
A.~J. Ross, L.~Samushia, C.~Howlett, W.~J. Percival, A.~Burden, and M.~Manera,
  ``The clustering of the sdss dr7 main galaxy sample--i. a 4 per cent distance
  measure at z= 0.15,'' {\em Monthly Notices of the Royal Astronomical
  Society}, vol.~449, no.~1, pp.~835--847, 2015.

\bibitem{6dfgs}
F.~Beutler, C.~Blake, M.~Colless, D.~H. Jones, L.~Staveley-Smith, G.~B. Poole,
  L.~Campbell, Q.~Parker, W.~Saunders, and F.~Watson, ``The 6df galaxy survey:
  $z=0$ measurements of the growth rate and $\sigma_8$,'' {\em Monthly Notices
  of the Royal Astronomical Society}, vol.~423, no.~4, pp.~3430--3444, 2012.

\bibitem{jin1996computation}
J.~Jin and Z.~S. Jjie, {\em Computation of special functions}.
\newblock Wiley, 1996.

\bibitem{camb}
C.~Howlett, A.~Lewis, A.~Hall, and A.~Challinor, ``Cmb power spectrum parameter
  degeneracies in the era of precision cosmology,'' {\em Journal of Cosmology
  and Astroparticle Physics}, vol.~2012, no.~04, p.~027, 2012.

\bibitem{maperteq}
C.-P. Ma and E.~Bertschinger, ``Cosmological perturbation theory in the
  synchronous and conformal newtonian gauges,'' {\em arXiv preprint
  astro-ph/9506072}, 1995.

\bibitem{cambnotes}
A.~Lewis, ``Camb notes,'' 2011.

\bibitem{Ichiki_2004}
K.~Ichiki, M.~Oguri, and K.~Takahashi, ``Constraints from the wilkinson
  microwave anisotropy probe on decayingcold dark matter,'' {\em Physical
  Review Letters}, vol.~93, Aug 2004.

\bibitem{classII}
D.~Blas, J.~Lesgourgues, and T.~Tram, ``The cosmic linear anisotropy solving
  system (class). part ii: approximation schemes,'' {\em Journal of Cosmology
  and Astroparticle Physics}, vol.~2011, no.~07, p.~034, 2011.

\bibitem{audren2014strongest}
B.~Audren, J.~Lesgourgues, G.~Mangano, P.~D. Serpico, and T.~Tram, ``Strongest
  model-independent bound on the lifetime of dark matter,'' {\em Journal of
  Cosmology and Astroparticle Physics}, vol.~2014, no.~12, p.~028, 2014.

\bibitem{heitmann2010coyote}
K.~Heitmann, M.~White, C.~Wagner, S.~Habib, and D.~Higdon, ``The coyote
  universe. i. precision determination of the nonlinear matter power
  spectrum,'' {\em The Astrophysical Journal}, vol.~715, no.~1, p.~104, 2010.

\bibitem{owls}
J.~Schaye, C.~D. Vecchia, C.~Booth, R.~P. Wiersma, T.~Theuns, M.~R. Haas,
  S.~Bertone, A.~R. Duffy, I.~McCarthy, and F.~van~de Voort, ``The physics
  driving the cosmic star formation history,'' {\em Monthly Notices of the
  Royal Astronomical Society}, vol.~402, no.~3, pp.~1536--1560, 2010.

\bibitem{mcalpine2016eagle}
S.~McAlpine, J.~C. Helly, M.~Schaller, J.~W. Trayford, Y.~Qu, M.~Furlong, R.~G.
  Bower, R.~A. Crain, J.~Schaye, T.~Theuns, {\em et~al.}, ``The eagle
  simulations of galaxy formation: Public release of halo and galaxy
  catalogues,'' {\em Astronomy and Computing}, vol.~15, pp.~72--89, 2016.

\bibitem{Smith:2002dz}
R.~E. Smith, J.~A. Peacock, A.~Jenkins, S.~D.~M. White, C.~S. Frenk, F.~R.
  Pearce, P.~A. Thomas, G.~Efstathiou, and H.~M.~P. Couchmann, ``{Stable
  clustering, the halo model and nonlinear cosmological power spectra},'' {\em
  Mon. Not. Roy. Astron. Soc.}, vol.~341, p.~1311, 2003.

\bibitem{Bird:2011rb}
S.~Bird, M.~Viel, and M.~G. Haehnelt, ``{Massive Neutrinos and the Non-linear
  Matter Power Spectrum},'' {\em Mon. Not. Roy. Astron. Soc.}, vol.~420,
  pp.~2551--2561, 2012.

\bibitem{takahashi}
R.~Takahashi, M.~Sato, T.~Nishimichi, A.~Taruya, and M.~Oguri, ``Revising the
  halofit model for the nonlinear matter power spectrum,'' {\em The
  Astrophysical Journal}, vol.~761, no.~2, p.~152, 2012.

\bibitem{hmcode}
A.~Mead, C.~Heymans, L.~Lombriser, J.~Peacock, O.~Steele, and H.~Winther,
  ``Accurate halo-model matter power spectra with dark energy, massive
  neutrinos and modified gravitational forces,'' {\em Monthly Notices of the
  Royal Astronomical Society}, vol.~459, no.~2, pp.~1468--1488, 2016.

\bibitem{dakin2019}
J.~Dakin, S.~Hannestad, and T.~Tram, ``Fully relativistic treatment of decaying
  cold dark matter in n-body simulations,'' {\em Journal of Cosmology and
  Astroparticle Physics}, vol.~2019, no.~06, p.~032, 2019.

\bibitem{cataneo2019road}
M.~Cataneo, L.~Lombriser, C.~Heymans, A.~Mead, A.~Barreira, S.~Bose, and B.~Li,
  ``On the road to percent accuracy: non-linear reaction of the matter power
  spectrum to dark energy and modified gravity,'' {\em Monthly Notices of the
  Royal Astronomical Society}, vol.~488, no.~2, pp.~2121--2142, 2019.

\bibitem{desy1ext}
T.~Abbott, F.~Abdalla, S.~Avila, M.~Banerji, E.~Baxter, K.~Bechtol, M.~Becker,
  E.~Bertin, J.~Blazek, S.~Bridle, {\em et~al.}, ``Dark energy survey year 1
  results: Constraints on extended cosmological models from galaxy clustering
  and weak lensing,'' {\em Physical Review D}, vol.~99, no.~12, p.~123505,
  2019.

\bibitem{Ade:2015rim}
P.~A.~R. Ade {\em et~al.}, ``{Planck 2015 results. XIV. Dark energy and
  modified gravity},'' {\em Astron. Astrophys.}, vol.~594, p.~A14, 2016.

\bibitem{planckpy}
H.~Prince and J.~Dunkley, ``Data compression in cosmology: A compressed
  likelihood for planck data,'' {\em Physical Review D}, vol.~100, no.~8,
  p.~083502, 2019.

\bibitem{cosmosis}
J.~Zuntz, M.~Paterno, E.~Jennings, D.~Rudd, A.~Manzotti, S.~Dodelson,
  S.~Bridle, S.~Sehrish, and J.~Kowalkowski, ``Cosmosis: Modular cosmological
  parameter estimation,'' {\em Astronomy and Computing}, vol.~12, pp.~45--59,
  2015.

\bibitem{kumar2018cosmological}
S.~Kumar, R.~C. Nunes, and S.~K. Yadav, ``Cosmological bounds on dark
  matter-photon coupling,'' {\em Physical Review D}, vol.~98, no.~4, p.~043521,
  2018.

\bibitem{redmagic}
R.~Cawthon, C.~Davis, M.~Gatti, P.~Vielzeuf, J.~Elvin-Poole, E.~Rozo,
  J.~Frieman, E.~S. Rykoff, A.~Alarcon, G.~M. Bernstein, {\em et~al.}, ``Dark
  energy survey year 1 results: calibration of redmagic redshift distributions
  in des and sdss from cross-correlations,'' {\em Monthly Notices of the Royal
  Astronomical Society}, vol.~481, no.~2, pp.~2427--2443, 2018.

\bibitem{metacalibration}
E.~S. Sheldon and E.~M. Huff, ``Practical weak-lensing shear measurement with
  metacalibration,'' {\em The Astrophysical Journal}, vol.~841, no.~1, p.~24,
  2017.

\bibitem{aghanim2019planck}
N.~Aghanim, Y.~Akrami, M.~Ashdown, J.~Aumont, C.~Baccigalupi, M.~Ballardini,
  A.~Banday, R.~Barreiro, N.~Bartolo, S.~Basak, {\em et~al.}, ``Planck 2018
  results. v. cmb power spectra and likelihoods,'' {\em arXiv preprint
  arXiv:1907.12875}, 2019.

\bibitem{taruya2014beyond}
A.~Taruya, K.~Koyama, T.~Hiramatsu, and A.~Oka, ``Beyond consistency test of
  gravity with redshift-space distortions at quasilinear scales,'' {\em
  Physical Review D}, vol.~89, no.~4, p.~043509, 2014.

\bibitem{barreira2016validating}
A.~Barreira, A.~G. Sanchez, and F.~Schmidt, ``Validating estimates of the
  growth rate of structure with modified gravity simulations,'' {\em Physical
  Review D}, vol.~94, no.~8, p.~084022, 2016.

\bibitem{handley2015polychord}
W.~Handley, M.~Hobson, and A.~Lasenby, ``Polychord: nested sampling for
  cosmology,'' {\em Monthly Notices of the Royal Astronomical Society:
  Letters}, vol.~450, no.~1, pp.~L61--L65, 2015.

\bibitem{multinest}
F.~Feroz, M.~P. Hobson, and M.~Bridges, ``{MultiNest: an efficient and robust
  Bayesian inference tool for cosmology and particle physics},'' {\em Monthly
  Notices of the Royal Astronomical Society}, vol.~398, pp.~1601--1614, 09
  2009.

\bibitem{benevento2020can}
G.~Benevento, W.~Hu, and M.~Raveri, ``Can late dark energy transitions raise
  the hubble constant?,'' {\em Physical Review D}, vol.~101, no.~10, p.~103517,
  2020.

\bibitem{raveri2019concordance}
M.~Raveri and W.~Hu, ``Concordance and discordance in cosmology,'' {\em
  Physical Review D}, vol.~99, no.~4, p.~043506, 2019.

\bibitem{handley2019}
W.~Handley and P.~Lemos, ``Quantifying tension: interpreting the des evidence
  ratio,'' {\em arXiv preprint arXiv:1902.04029}, 2019.

\bibitem{wu2020hubble}
W.~Wu, P.~Motloch, W.~Hu, and M.~Raveri, ``Hubble constant tension between cmb
  lensing and bao measurements,'' {\em arXiv preprint arXiv:2004.10207}, 2020.

\bibitem{anesthetic}
W.~Handley, ``anesthetic: nested sampling visualisation,'' {\em The Journal of
  Open Source Software}, vol.~4, p.~1414, Jun 2019.

\bibitem{knuth2015bayesian}
K.~H. Knuth, M.~Habeck, N.~K. Malakar, A.~M. Mubeen, and B.~Placek, ``Bayesian
  evidence and model selection,'' {\em Digital Signal Processing}, vol.~47,
  pp.~50--67, 2015.

\bibitem{jeffreys1961theory}
H.~Jeffreys, ``Theory of probability, clarendon,'' 1961.

\bibitem{robert2009harold}
C.~P. Robert, N.~Chopin, J.~Rousseau, {\em et~al.}, ``Harold jeffreys’s
  theory of probability revisited,'' {\em Statistical Science}, vol.~24, no.~2,
  pp.~141--172, 2009.

\bibitem{riess2019large}
A.~G. Riess, S.~Casertano, W.~Yuan, L.~M. Macri, and D.~Scolnic, ``Large
  magellanic cloud cepheid standards provide a 1\% foundation for the
  determination of the hubble constant and stronger evidence for physics beyond
  $\lambda$cdm,'' {\em The Astrophysical Journal}, vol.~876, no.~1, p.~85,
  2019.

\bibitem{giblin2019road}
B.~Giblin, M.~Cataneo, B.~Moews, and C.~Heymans, ``On the road to per cent
  accuracy--ii. calibration of the non-linear matter power spectrum for
  arbitrary cosmologies,'' {\em Monthly Notices of the Royal Astronomical
  Society}, vol.~490, no.~4, pp.~4826--4840, 2019.

\bibitem{d2020cosmological}
G.~d'Amico, J.~Gleyzes, N.~Kokron, K.~Markovic, L.~Senatore, P.~Zhang,
  F.~Beutler, and H.~Gil-Mar{\'\i}n, ``The cosmological analysis of the
  sdss/boss data from the effective field theory of large-scale structure,''
  {\em Journal of Cosmology and Astroparticle Physics}, vol.~2020, no.~05,
  p.~005, 2020.

\end{thebibliography}

\begin{figure*}
\centering
 \includegraphics[width=0.8\textwidth]{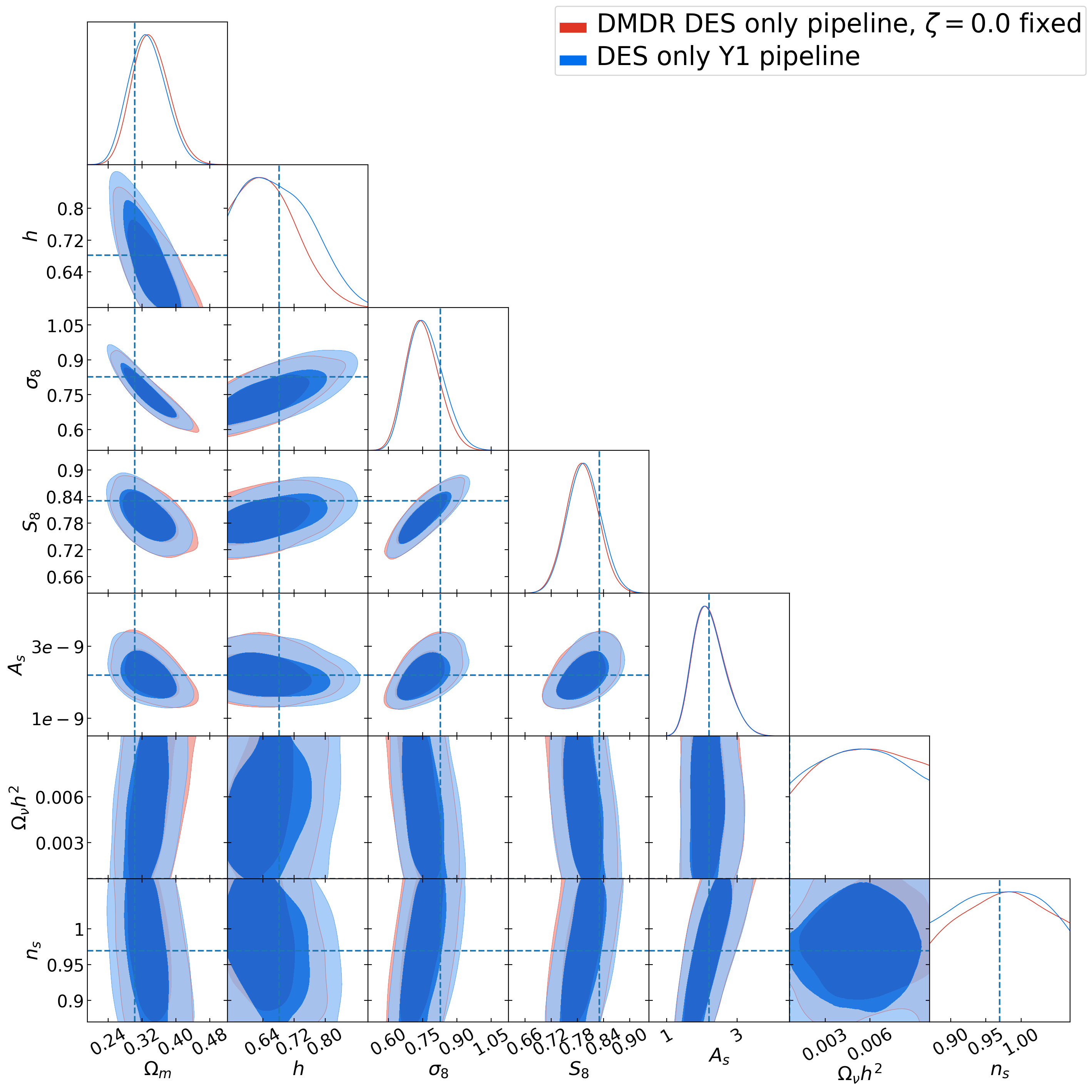}
 \caption{Comparison of the constraints using DES-Y1 analysis pipeline (blue) and our DMDR analysis pipeline with new parameters fixed ($\zeta=0.0$, $\kappa=1.0$; red contours). We use a simulated  $\Lambda$CDM data vector on which we apply the  {\fontfamily{qcr}\selectfont multinest} MCMC chains for both runs. }
\label{fig:pipelined}
\end{figure*}

\begin{figure*}
\centering
\includegraphics[width=0.8\textwidth]{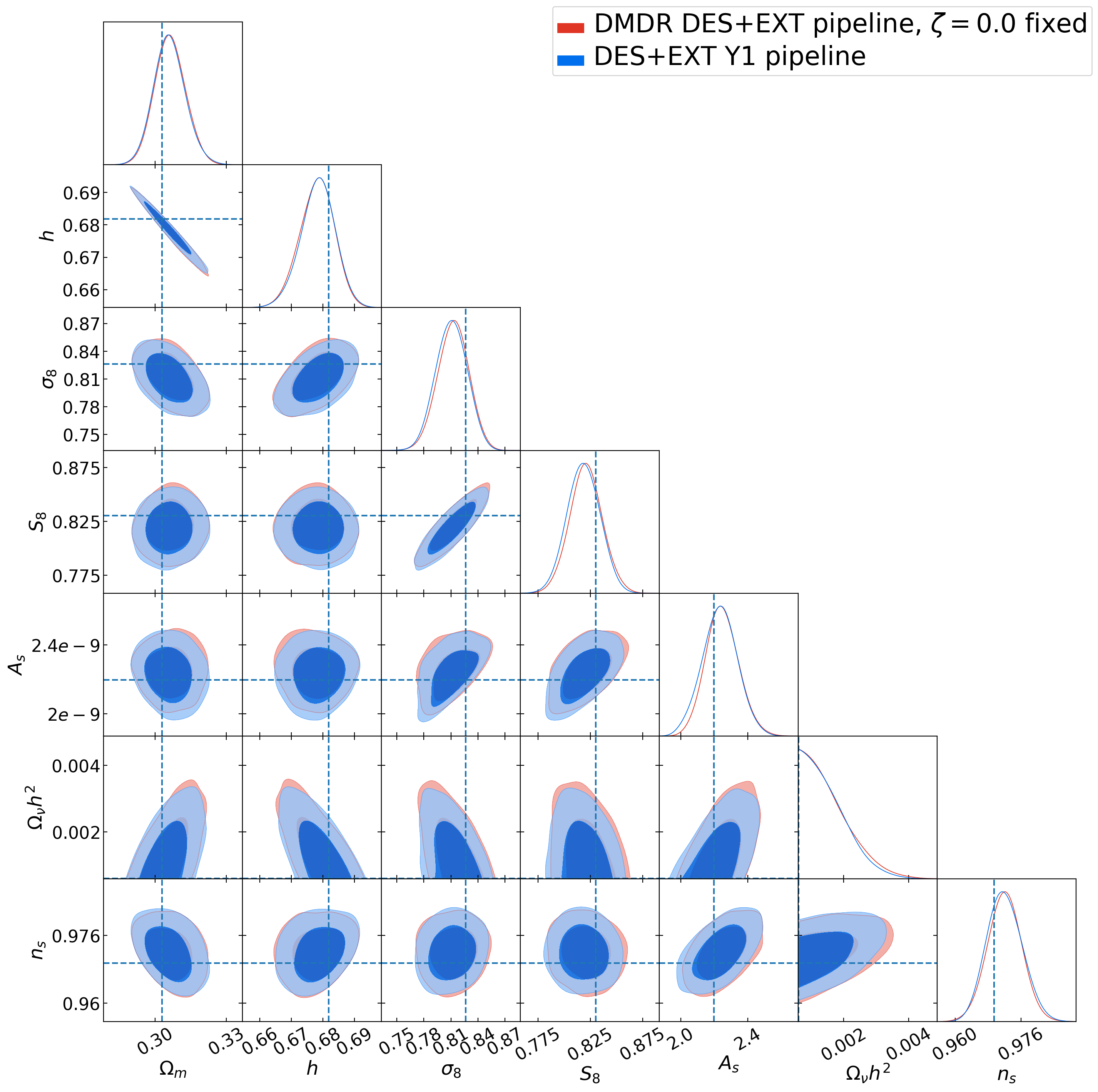}
\caption{Same as figure~\ref{fig:pipelined}, but for DES-Y1+External simulated data.}
\label{fig:pipelinede}
\end{figure*}

\appendix 

\section{Pipeline Comparison on $\Lambda$CDM}
\label{sec:pipelinecomparison}

We want to make sure that, any cosmological parameters constraints that are found different from the DES-Y1 3x2pt Key paper \cite{desy1} ones are physical, namely caused by the DMDR model, but not due to the pipeline choices variance. Hence we run full {\fontfamily{qcr}\selectfont
multinest} MCMC chains on the same $\Lambda$CDM simulated data vector, using DES-Y1 analysis pipeline and our DMDR analysis pipeline with $\zeta=0.0$, $\kappa=1.0$ fixed ($\Lambda$CDM subspace, so $\kappa$ value is irrelevant). The results are shown in figures~\ref{fig:pipelined} and \ref{fig:pipelinede} for DES only and DES+External Data. In both cases, except for the parameters that are not effectively constrained like $h$, $\Omega_{\nu}h^2$ and $n_s$ for DES only data, the posteriors from two pipelines agree with each other at $\lesssim 0.1 \sigma$ level.

\section{Dark Radiation Hierarchy equations}
\label{app:perturbation}

In B18, perturbation equations were derived from the perturbation expansion of the energy--momentum tensor for dark matter and dark radiation,
\begin{align}
    T^{\rm dm}_{\mu \nu} &= \bar{\rho}_{\rm dm}(1+\delta_{\rm dm})u_{\mu}^{\rm dm} u_{\nu}^{\rm dm}\\
    T^{\rm dr}_{\mu \nu}  &= \frac{4}{3}\bar{\rho}_{\rm dr}(1+\delta_{\rm dr}) u_{\mu}^{\rm dr}u_{\nu}^{\rm dr} + \frac{\bar{\rho}_{\rm dr}(1+\delta_{\rm dr})}{3}g_{\mu\nu}+\Pi^{\rm dr}_{\mu \nu}
\end{align}
where in synchronous gauge $u_{\mu}^{\rm dm} = a(1,\vec{0})$, $u_{\mu}^{\rm dr} = a(1,\vec{v}^{\rm dr})$. For dark matter and dark radiation defined in this way, we can write the continuity equations and Einstein equations as:
\begin{align}
\nabla^{\nu}T_{\mu \nu}^{\rm dm} & =   -\nabla^{\nu}T_{\mu \nu}^{\rm dr} = -\mathcal{Q}u_{\mu}^{\rm dm}\\
R_{\mu\nu}-\frac{1}{2}R g_{\mu\nu}+\Lambda g_{\mu\nu} & =  \frac{8\pi G}{c^4} T_{\mu \nu}
\end{align}
where $u_{\mu}^{\rm dm}$ is the proper velocity of the dark matter. Note that the right-hand side of the continuity equation has a collision term instead of zero for CDM. In B18 the dark radiation is only expanded up to $\delta_{\rm dr}$, $\theta_{\rm dr}=\partial_i v_{\rm dr}^i$ and one anisotropy shear $\Pi^{\rm dr}_{ij} = (\partial_i \partial_j - \frac{1}{3}\delta_{ij}\nabla^2)\Pi^{\rm dr}$, which is sufficient when dark radiation self-interacts or continues to interact with dark matter after produced so the higher $\ell$ terms damp out. 

In our work, we assume dark radiation to be a completely free-streaming relativistic species and write down the full phase space perturbation hierarchy equations for it, which differs from the massless neutrino ones by a collision term. The phase space dynamics of the dark radiation with collision terms are \cite{maperteq}:
\begin{equation}
    \begin{split}
    \frac{\partial F_{\rm dr}(\vec{k},\hat{n},\tau)}{\partial \tau} & +  ik\mu F_{\rm dr}(\vec{k},\hat{n},\tau) = -\frac{2}{3}\dot{h}(\vec{k},\tau) - \frac{4}{3}(\dot{h}(\vec{k},\tau) \\ 
    & +6\dot{\eta}(\vec{k},\tau))P_2(\hat{k}\cdot \hat{n}) + \left( \frac{\partial F_{\rm dr}(\vec{k},\hat{n},\tau)}{\partial \tau}\right)_C\\
    \end{split}
\end{equation}
The phenomenology of the microphysics of the dark matter to dark radiation conversion process is mostly demonstrated in the collision term 
\begin{equation}
    \left( \frac{\partial F_{\rm dr}(\vec{k},\hat{n},\tau)}{\partial \tau}\right)_C = \frac{a}{\rho_{\rm dr}(a)}(-\mathcal{Q}(a)F_{\rm dr}(\vec{k},\hat{n},\tau)+\delta \mathcal{Q}),
\end{equation}
especially its perturbation part $\delta \mathcal{Q}$ which depends on the details of the interacting physical quantities like particle momentum.
However, from several case studies in B18 on Sommerfeld enhancement and single-body decay processes, it seems that the precision of the current generation of cosmological observations is not sufficient to discriminate between the specific forms of $\delta \mathcal{Q}$. Hence we assume the simplest form of the collision perturbation $\delta \mathcal{Q} = \mathcal{Q} \delta_{\rm dm}$, without dependence on polarization or momentum anisotropy: 
\begin{equation}
    \left( \frac{\partial F_{\rm dr}(\vec{k},\hat{n},\tau)}{\partial \tau}\right)_C = \frac{\mathcal{Q}(a)a}{\rho_{\rm dr}(a)} (-F_{\rm dr}(\vec{k},\hat{n},\tau)+\delta_{\rm dm}(\vec{k},\tau))
    \label{eq:appcollision}
\end{equation}

Expanding $F_{\rm dr}$ in equation~(\ref{eq:appcollision}) into harmonics, we get
\begin{equation}
    F_{\rm dr}(\vec{k},\hat{n},\tau) = \sum_{l=1}^{\infty}(-i)^l (2l+1)F_{\rm dr}l(\vec{k},\tau)P_l(\hat{k}\cdot\hat{n}).
\end{equation}
Noticing that only $F_{\rm dr}(\vec{k},\hat{n},\tau)$ itself needs expansion while other terms in equation~(\ref{eq:appcollision}) are constant to the orientation variable $\hat{k}\cdot \hat{n}$, we get the hierarchy equation 
\cite{maperteq,cambnotes,classII,audren2014strongest}: 
\begin{equation}
\begin{split}
(J^{\rm dr}_l)' &=  \frac{k}{2l+1}
[lJ_{l-1}^{\rm dr}-\beta_{l+1}(l+1)J_{l+1}^{\rm dr}] \\
&+ \frac{8}{15}k\sigma \delta_{l2}  -\frac{4}{3} k \mathcal{Z} \delta_{l0} - \frac{a Q}{\bar{\rho}_{\rm dr}}J_l^{\rm dr}
\end{split}
\label{appeq:hierarchy}
\end{equation}
where $J^{\rm dr}_0 \equiv \delta_{\rm dr}$, $J^{\rm dr}_1 \equiv q_{\rm dr} = \frac{4}{3}\theta_{\rm dr}/k$, $J^{\rm dr}_2 \equiv \pi_{\rm dr}=\Pi^{\rm dr}/\bar{\rho}_{\rm dr}$ in CAMB convention, $\delta_{l0},\delta_{l2}$ are Dirac delta-functions. Equations $l=0, l=1$ agree with the Eqs.~(14) and (15) in B18.

\clearpage
\end{document}